\documentclass[twoside,twocolumn,9pt]{article}
\usepackage{extsizes}
\usepackage[super,sort&compress,comma]{natbib} 
\usepackage[version=3]{mhchem}
\usepackage[left=1.5cm, right=1.5cm, top=1.785cm, bottom=2.0cm]{geometry}
\usepackage{balance}
\usepackage{mathptmx}
\usepackage{sectsty}
\usepackage{graphicx} 
\usepackage{lastpage}
\usepackage[format=plain,justification=justified,singlelinecheck=false,font={stretch=1.125,small,sf},labelfont=bf,labelsep=space]{caption}
\usepackage{float}
\usepackage{fancyhdr}
\usepackage{fnpos}
\usepackage[english]{babel}
\addto{\captionsenglish}{%
  \renewcommand{\refname}{Notes and references}
}
\usepackage{array}
\usepackage{droidsans}
\usepackage{charter}
\usepackage[T1]{fontenc}
\usepackage[usenames,dvipsnames]{xcolor}
\usepackage{setspace}
\usepackage[compact]{titlesec}
\usepackage{hyperref}
\usepackage{epstopdf}%This line makes .eps figures into .pdf - please comment out if not required.

\definecolor{cream}{RGB}{222,217,201}
\DeclareUnicodeCharacter{FFFC}{}

\begin{document}

\pagestyle{fancy}
\thispagestyle{plain}
\fancypagestyle{plain}{
%%%HEADER%%%
\renewcommand{\headrulewidth}{0pt}
}
%%%END OF HEADER%%%

%%%PAGE SETUP - Please do not change any commands within this section%%%
\makeFNbottom
\makeatletter
\renewcommand\LARGE{\@setfontsize\LARGE{15pt}{17}}
\renewcommand\Large{\@setfontsize\Large{12pt}{14}}
\renewcommand\large{\@setfontsize\large{10pt}{12}}
\renewcommand\footnotesize{\@setfontsize\footnotesize{7pt}{10}}
\makeatother

\renewcommand{\thefootnote}{\fnsymbol{footnote}}
\renewcommand\footnoterule{\vspace*{1pt}% 
\color{cream}\hrule width 3.5in height 0.4pt \color{black}\vspace*{5pt}} 
\setcounter{secnumdepth}{5}

\makeatletter 
\renewcommand\@biblabel[1]{#1}            
\renewcommand\@makefntext[1]% 
{\noindent\makebox[0pt][r]{\@thefnmark\,}#1}
\makeatother 
\renewcommand{\figurename}{\small{Fig.}~}
\sectionfont{\sffamily\Large}
\subsectionfont{\normalsize}
\subsubsectionfont{\bf}
\setstretch{1.125} %In particular, please do not alter this line.
\setlength{\skip\footins}{0.8cm}
\setlength{\footnotesep}{0.25cm}
\setlength{\jot}{10pt}
\titlespacing*{\section}{0pt}{4pt}{4pt}
\titlespacing*{\subsection}{0pt}{15pt}{1pt}
%%%END OF PAGE SETUP%%%

%%%FOOTER%%%
\fancyfoot{}
\fancyfoot[LO,RE]{\vspace{-7.1pt}\includegraphics[height=9pt]{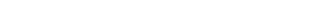}}
\fancyfoot[CO]{\vspace{-7.1pt}\hspace{13.2cm}\includegraphics{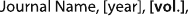}}
\fancyfoot[CE]{\vspace{-7.2pt}\hspace{-14.2cm}\includegraphics{head_foot/RF}}
\fancyfoot[RO]{\footnotesize{\sffamily{1--\pageref{LastPage} ~\textbar  \hspace{2pt}\thepage}}}
\fancyfoot[LE]{\footnotesize{\sffamily{\thepage~\textbar\hspace{3.45cm} 1--\pageref{LastPage}}}}
\fancyhead{}
\renewcommand{\headrulewidth}{0pt} 
\renewcommand{\footrulewidth}{0pt}
\setlength{\arrayrulewidth}{1pt}
\setlength{\columnsep}{6.5mm}
\setlength\bibsep{1pt}
%%%END OF FOOTER%%%

%%%FIGURE SETUP - please do not change any commands within this section%%%
\makeatletter 
\newlength{\figrulesep} 
\setlength{\figrulesep}{0.5\textfloatsep} 

\newcommand{\topfigrule}{\vspace*{-1pt}% 
\noindent{\color{cream}\rule[-\figrulesep]{\columnwidth}{1.5pt}} }

\newcommand{\botfigrule}{\vspace*{-2pt}% 
\noindent{\color{cream}\rule[\figrulesep]{\columnwidth}{1.5pt}} }

\newcommand{\dblfigrule}{\vspace*{-1pt}% 
\noindent{\color{cream}\rule[-\figrulesep]{\textwidth}{1.5pt}} }

\makeatother
%%%END OF FIGURE SETUP%%%

%%%TITLE, AUTHORS AND ABSTRACT%%%
\twocolumn[
  \begin{@twocolumnfalse}
{\includegraphics[height=30pt]{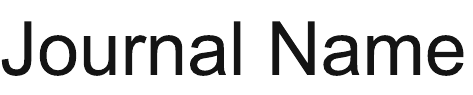}\hfill\raisebox{0pt}[0pt][0pt]{\includegraphics[height=55pt]{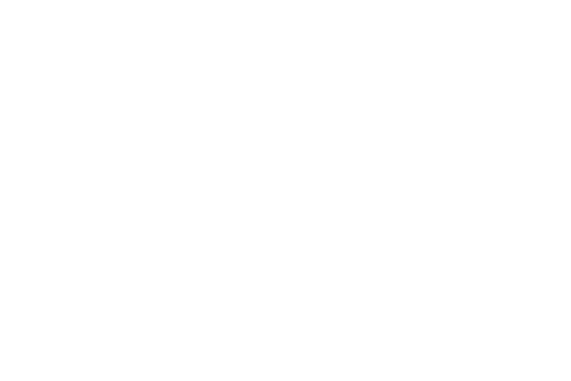}}\\[1ex]
\includegraphics[width=18.5cm]{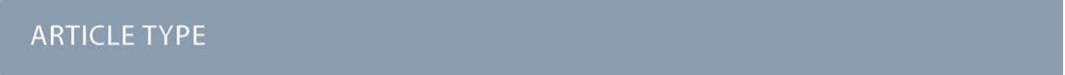}}\par
\vspace{1em}
\sffamily
\begin{tabular}{m{4.5cm} p{13.5cm} }

\includegraphics{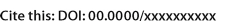} & \noindent\LARGE{\textbf{\ce{CO2} Adsorption Mechanisms in Hydrated Silica Nanopores: Insights From Grand Canonical Monte Carlo to Classical and Ab Initio Molecular Dynamics
$^\dag$}} \\%Article title goes here instead of the text "This is the title"
\vspace{0.3cm} & \vspace{0.3cm} \\

 & \noindent\large{Jihong Shi,\textit{$^{a}$} Tao Zhang,$^{\ast}$\textit{$^{b}$}
 %textit{$^{b\ddag}$} 
 Shuyu Sun,\textit{$^{c}$} and Liang Gong$^{\ast}$\textit{$^{b}$}} \\%Author names go here instead of "Full name", etc.

\includegraphics{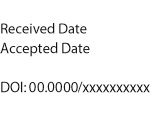} & \noindent\normalsize{Understanding interfacial phenomena in confined systems is important for optimizing \ce{CO2} capture technologies. Here, we present a comprehensive investigation of \ce{CO2} adsorption in hydrated amorphous silica nanopores through an integrated computational approach combining grand canonical Monte Carlo (GCMC), classical molecular dynamics (MD), and ab initio molecular dynamics (AIMD) simulations. The excess adsorption isotherms reveal a marked hydration dependence, with \ce{CO2} uptake decreasing from 7.6 to 2.6 mmol/g as water content increases from 1 to 15 wt\%. Analysis of adsorption kinetics demonstrates a distinctive bimodal process, characterized by rapid initial uptake followed by slower diffusion-limited adsorption, with the latter becoming increasingly dominant at higher hydration levels. Classical MD simulations reveal an inverse correlation between hydration and \ce{CO2} mobility, with self-diffusion coefficients decreasing 
%from $2.8\times10^{-9}$ to $0.9\times10^{-9}$ m$^2$/s 
across the studied hydration range. Density profile analysis indicates a hydration-induced transition in \ce{CO2} distribution from central pore regions to surface-proximate domains, accompanied by restructuring of interfacial water networks. Notably, AIMD simulations capture previously unrecognized chemical processes, including proton transfer mechanisms leading to surface silanol formation with characteristic O-O distances of 2.4-2.5 Å, and spontaneous \ce{CO2} hydration yielding carbonate species through water-mediated reaction pathways. These findings demonstrate the dual role of confined water as both a spatial competitor and reaction medium for \ce{CO2} capture, providing molecular-level insights with quantum mechanical accuracy for design of carbon capture materials.} \\%The abstrast goes here instead of the text "The abstract should be..."

\end{tabular}

 \end{@twocolumnfalse} \vspace{0.6cm}

  ]
%%%END OF TITLE, AUTHORS AND ABSTRACT%%%

%%%FONT SETUP - please do not change any commands within this section
\renewcommand*\rmdefault{bch}\normalfont\upshape
\rmfamily
\section*{}
\vspace{-1cm}

%%%FOOTNOTES%%%

\footnotetext{\textit{$^{a}$~Department of Physics, King's College London, Strand, London WC2R 2LS, UK.}}
\footnotetext{\textit{$^{b}$~College of New Energy, China University of Petroleum (East China), Qingdao 266580, China. ${\ast}$E-mail: tao.zhang@upc.edu.cn, lgong@upc.edu.cn.}}
\footnotetext{\textit{$^{c}$~Computational Mathematics, School of Mathematical Sciences, Tongji University, Shanghai 200092, China.}}
%footnotetext{\textit{$^{c}$~Computational Transport Phenomena Laboratory, Division of Physical Science and Engineering, King Abdullah University of Science and Technology, Thuwal 23955-6900, Saudi Arabia.}}
%Please use \dag to cite the ESI in the main text of the article.
%If you article does not have ESI please remove the the \dag symbol from the title and the footnotetext below.
\footnotetext{\dag~Electronic Supplementary Information (ESI) available. See DOI: xxx.}

%%%END OF FOOTNOTES%%%

%%%MAIN TEXT%%%%
\section{Introduction}
The accelerating accumulation of atmospheric \ce{CO2} has emerged as a critical driver of global climate change\cite{bouwer2013projections,masson2021climate,boubaker2024carbon}. Carbon capture, utilization, and storage (CCUS) technologies have emerged as key strategies for reducing atmospheric \ce{CO2} levels and meeting carbon neutrality goals.\cite{gabrielli2020role}. Among various CCUS approaches, \ce{CO2} geological sequestration\cite{lal2008carbon}, which involves the injection of captured \ce{CO2} into underground formations such as depleted oil and gas reservoirs\cite{gbadamosi2019overview}, deep saline aquifers\cite{10.1007/978-981-97-0268-8_34,yang2010characteristics}, and unmineable coal seams\cite{talapatra2020study}, is considered a promising long-term solution for reducing atmospheric \ce{CO2} levels\cite{en17195000}. Simultaneously, the development of advanced materials for surface-based \ce{CO2} capture is being actively pursued, focusing on porous structures like silica aerogels\cite{akhter2021silica}, metal-organic frameworks\cite{pettinari2020metal}, and hybrid composites\cite{navik2024atmospheric,ramasamy2024leveraging} that can selectively adsorb \ce{CO2} from industrial flue gases. 
%Both strategies aim to support the transition to a low-carbon economy by offering effective pathways for \ce{CO2} storage and capture.

Amorphous silica, a major component of various geological formations and a versatile synthetic material\cite{chittick1969preparation}, is of particular interest for both sub-surface and surface-based \ce{CO2} capture applications due to its high surface area and chemical reactivity \cite{dias2022silica}. Understanding the underlying mechanisms of \ce{CO2} adsorption and diffusion in amorphous silica is crucial to optimizing its use as a sequestration medium and enhancing its efficiency for long-term \ce{CO2} storage. Despite extensive research in this domain, significant gaps remain in correlating the microscopic adsorption capacity of \ce{CO2} with the electronic-level interactions that drive adsorption, particularly under realistic geological conditions. This study aims to address these gaps by utilizing an integrated methodology that combines grand canonical Monte Carlo (GCMC)\cite{doi:10.1080/00268977500100221}, classical molecular dynamics (MD), and ab initio molecular dynamics (AIMD) to systematically examine the adsorption dynamics, diffusion capacity, and molecular interactions of \ce{CO2} in an amorphous silica slit model under hydrated condition.

Experimental studies have played an important role in validating simulation results and providing macroscopic data on \ce{CO2} adsorption capacity and selectivity. \citeauthor{khoshraftar2021evaluation}\cite{khoshraftar2021evaluation} studied the adsorption properties of \ce{CO2} on silica gel and evaluated its potential as a low-cost adsorbent for \ce{CO2} capture. This study analyzed the \ce{CO2} adsorption capacity at different pressures (2 to 8 bar) and temperatures ($25^{\circ}\mathrm{C}$ to $85^{\circ}\mathrm{C}$) through experiments and characterized the silica gel using techniques such as XRD, BET and FTIR. They found that \ce{CO2} adsorption increased with increasing pressure and decreased with increasing temperature, indicating that the adsorption process is exothermic\cite{khoshraftar2021evaluation}. The adsorption characteristics of \ce{CO2} on mesoporous silica and titania surfaces functionalized with aminopropylsilane (APS) were investigated by \citeauthor{doi:10.1021/jp907054h} combining microcalorimetry and in situ FTIR spectroscopy \cite{doi:10.1021/jp907054h}. The APS functionalization introduces amino groups on silica and titania surfaces, intended to improve \ce{CO2} adsorption efficiency. The detailed differentiation between carbamate formation on silica and bidentate carbonate formation on titania highlights the importance of material selection for \ce{CO2} capture applications. \citeauthor{doi:10.1021/la010061q}\cite{doi:10.1021/la010061q} used a Rubotherm magnetic suspension balance with in situ density measurement to quantify carbon dioxide isotherms on silica at various temperatures (311 - 466 K) and pressures (0 - 450 bar). This approach measures the excess adsorbed quantity and estimates adsorption heat without making assumptions about the adsorbed phase. They found that \ce{CO2} adsorption increased with density, reaching a maximum before declining, indicating that density is a more effective variable than pressure for describing high-pressure adsorption. Adsorption enthalpies also increased with surface potential, especially at lower temperatures, highlighting enhanced adsorption near the critical temperature of \ce{CO2}. While the study achieved high precision, limitations include challenges in measuring near the critical point.

Theoretical studies employing molecular simulations \cite{shi2024characterizing,doi:10.1021/acs.langmuir.4c03177,liu2024thermal,khosrowshahi2022role,GONG2020118406,C9RA04963K} have significantly advanced our understanding of \ce{CO2} adsorption behaviour at the atomic scale in various porous media. \citeauthor{doi:10.1021/acs.energyfuels.2c03244}\cite{doi:10.1021/acs.energyfuels.2c03244} investigated the adsorption behaviour of shale gas (the main component of which is methane, $\mathrm{CH}_4$) in kerogen matrix, under water-bearing and \ce{CO2}-containing conditions using MD simulations.%By analyzing the spatial density distribution, free energy and potential energy of different molecules in slit nanopores, 
They found that the adsorption behaviour of each component showed significant inhomogeneity within the organic matrix. Moisture favours adsorption at oxygen-rich sites due to strong polarity and hydrogen bonding interactions, whereas \ce{CO2} tends to adsorb at carbon-rich sites, which is mainly influenced by van der Waals forces and molecular polarity. The interactions between \ce{CO2}, $\mathrm{H}_2\mathrm{O}$, and kerogen are heavily influenced by induced polarization effects.
%, especially in the presence of water, which is highly polarizable.
However, the pcff+ force field\cite{yiannourakou2013molecular} employed in their work\cite{doi:10.1021/acs.energyfuels.2c03244} relies on fixed atomic charges, limiting its ability to capture the polarization effects that arise from molecular adsorption. To investigate the effects of surface disorder and defects, \citeauthor{TURCHI2024122709}\cite{TURCHI2024122709} investigated \ce{CO2} adsorption in amorphous nanopores versus crystalline surfaces using the ClayFF force field\cite{doi:10.1021/jp0363287}. While their study revealed the impact of surface heterogeneity and coordination defects, the classical force field approach limits accurate description of defect-related charge distributions. More sophisticated methods such as AIMD or machine learning potentials\cite{wang2018deepmd} would be necessary to properly capture these electronic effects. \citeauthor{sui2024competitive}\cite{sui2024competitive} investigated the competitive adsorption behaviour of \ce{CO2} and $\mathrm{CH}_4$ on functionalized silica surfaces using GCMC and MD simulations, specifically focusing on the impact of different functional groups (e.g., $\mathrm{COOH}$, $\mathrm{NH}_2$) on adsorption selectivity. 
%The work utilizes both GCMC and MD simulations to calculate adsorption isotherms, heat of adsorption, and self-diffusion coefficients across various temperatures and pressures. 
Critically, the study provides valuable insights into how specific functional groups enhance \ce{CO2} selectivity through strong electrostatic interactions, particularly the $\mathrm{COOH}$ group. While these findings provide valuable insights, the use of polarizable force fields or AIMD would better capture charge redistribution effects, particularly for systems containing both polar and nonpolar species. \citeauthor{deng2024mineralization}\cite{deng2024mineralization} employed a multi-scale approach combining GCMC, MD, and density functional theory (DFT) to investigate \ce{CO2} adsorption and mineralization mechanisms in feldspar composites, focusing on carbonate formation in \ce{K^+}-rich regions. Their DFT calculations, however, did not incorporate dispersion corrections, potentially underestimating the long-range van der Waals interactions between gas molecules and the substrate.

Despite the significant advancements made in both simulation and experimental research, several critical limitations hinder a comprehensive understanding of \ce{CO2} behaviour in amorphous silica. One major challenge is the discrepancy between idealized simulation models and the heterogeneous nature of real geological materials. Most simulations assume uniform pore structures and surface chemistries, which may not accurately capture the defect-rich and chemically diverse surfaces of natural amorphous silica\cite{STALLONS20014205}. Experimentally, replicating the complex dynamic nature of real reservoirs and accurately characterizing the temporal evolution of \ce{CO2} adsorption and desorption processes remain formidable. Moreover, although there are currently some studies on \ce{CO2} adsorption under hydrated conditions\cite{doi:10.1021/acs.langmuir.4c02136,liu2024thermal,GONG2020118406,long2021adsorption,liu2013molecular}, most of these investigations qualitatively describe how the presence of water inhibits \ce{CO2} adsorption capacity. However, due to the limitations of classical MD simulations using the empirical force fields, providing a detailed description of the interactions between water molecules and the substrate at the electronic level is challenging. These limitations make it difficult to fully understand the molecular mechanisms governing the effect of water content on \ce{CO2} adsorption. These fundamental challenges motivate the development of more sophisticated approaches to describe \ce{CO2}-silica interactions under realistic conditions.

To tackle these research challenges, this study utilizes an integrated simulation approach combining GCMC, MD, and AIMD techniques, each offering unique insights into different aspects of \ce{CO2} behaviour. GCMC is employed to predict equilibrium adsorption isotherms and evaluate the selectivity and capacity of \ce{CO2} under varying thermodynamic and water content conditions. Classical MD simulations are used to explore the dynamics of \ce{CO2} diffusion and adsorption at the atomic level, capturing the influence of moisture on \ce{CO2} mobility and interaction with local surface sites. AIMD simulations, leveraging quantum mechanical accuracy, are applied to investigate the electronic structure and reaction dynamics at defect sites, elucidating potential pathways for $\mathrm{H}_2\mathrm{O}$ and \ce{CO2} chemisorption and mineralization.

\section{Methodology}
\subsection{Amorphous Silica Slit Pore Model}
\begin{figure}[h]
\centering
  \includegraphics[height=7.3cm]{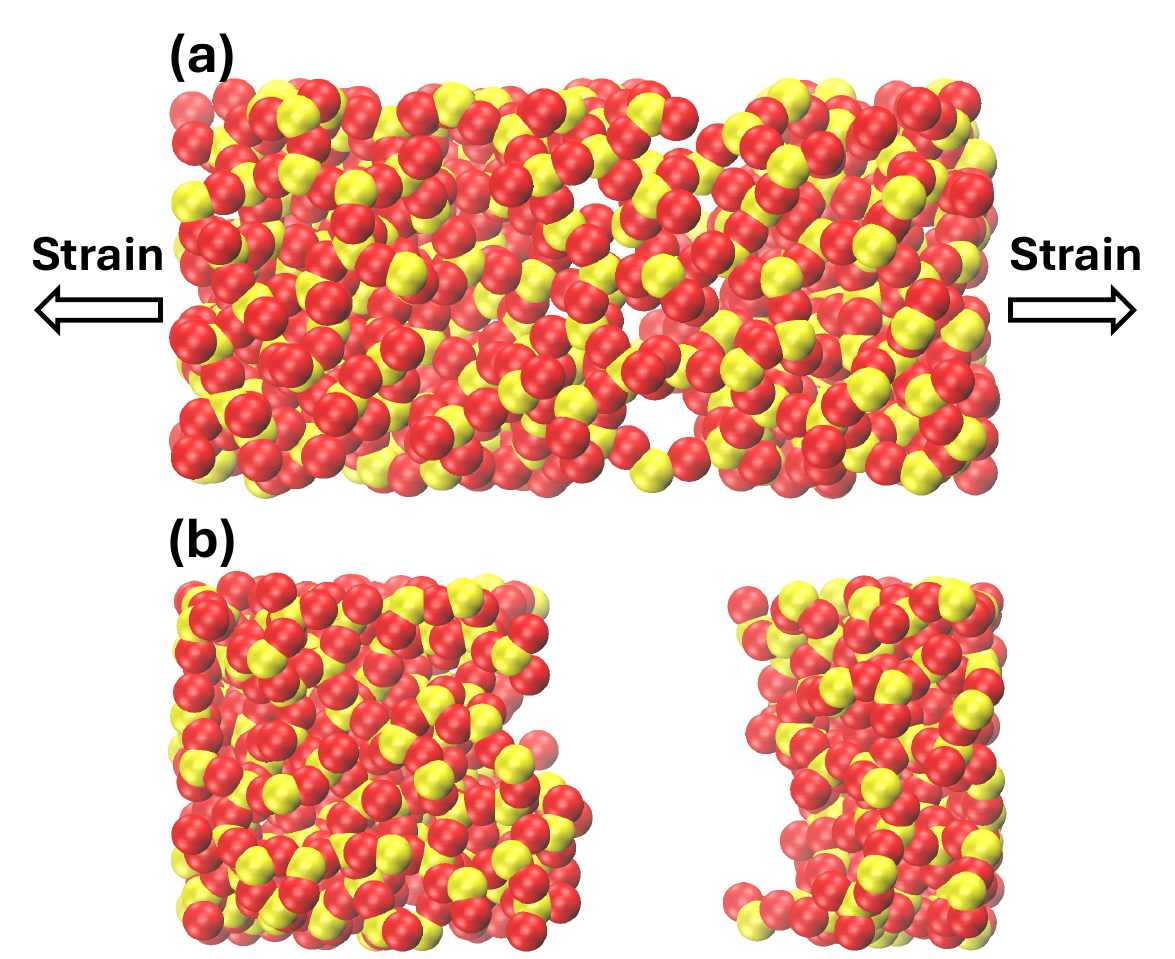}
  \caption{(a) Amorphous silica block strained in the z-direction. (b) The eventual amorphous silica natural slit model, corresponding to the average slit width approximately at 15 \AA. $\mathrm{Si}$ atoms are coloured in yellow, and oxygen atoms are red.}
  \label{fgr:model}
\end{figure}

Molecular dynamics simulations were performed for silica ($\mathrm{SiO}_2$) using the Vashishta potential\cite{PhysRevB.41.12197} that derives the bonds based on the relative locations of the atoms. Given this bond-angle energy-based potential, we do not have to provide bond and angle information in the configuration file as we do with the AMBER force field\cite{doi:10.1021/cm500365c}. The amorphous $\mathrm{SiO}_2$ block containing 864 atoms was built with the dimension of $19.593 \times 19.749 \times 29.570$ \AA$^3$ by replicating the ordered silica unit cell containing 3 Si and 6 O atoms and annealing the system. The entire annealing process is a finely tuned temperature and pressure process using the NPT system. These simulations using the Vashishta force field\cite{PhysRevB.41.12197}  need a smaller time step of 0.001 ps in \textit{metal} unit, which enables the modelling of bond formation and breaking. This is necessary to generate amorphous silica. The time step was 0.001 ps for these simulations. Firstly, the whole system runs 60,000 steps to ensure the melting of the $\mathrm{SiO}_2$ structure at 2000 K and keeps the pressure of the system equal to 100 atm in all three directions. Then in the second step, the system is run for another 60,000 steps and the temperature is lowered from 2000 K to 1000 K, but the isotropic barostat is changed to an anisotropic one and maintained at 100 atm. The third step is to continue to reduce the temperature of the system from 1000 K to our target temperature of 300 K, the anisotropic pressure barostat from the initial 100 atm down to the last 1 atm, run 300,000 steps, the fourth step, is to keep the temperature of the system unchanged at 300 K, to keep the initial and end of the pressure barostat are at 1 atm, run 60,000 steps. The initial model of the constructed amorphous silica block is shown in Fig.~\ref{fgr:model}a. The simulations are carried out using the large-scale atomic/molecular massively parallel simulator (LAMMPS, Nov 21, 2023 version) package.\cite{plimpton1995fast} 
%The cutoff for the long-range Coulomb and Van der Waals interactions is set as 12 \AA. A time step of 1 fs is used. 
Temperature and pressure are controlled with a Nos\'e-Hoover thermostat and barostat with relaxation times of 0.1 ps and 1 ps, respectively.

Once the amorphous silica block is obtained, it is then a matter of deforming the block and automatically cracking it into the natural slit model we need. The specific steps are as follows. In this step of MD simulation, we will use the ReaxFF\cite{van2001reaxff} force field, as the previous Vashishta force field\cite{PhysRevB.41.12197} is commonly used to describe more stable $\mathrm{SiO}_2$-bonded structures. This force field is suitable for describing the conventional $\mathrm{SiO}_2$ network structure, especially in the absence of chemical reactions. However, the reactive force field ReaxFF\cite{van2001reaxff} is able to dynamically adjust the strength and shape of the bonds to more accurately model bond breaking and chemical reactions and is ideal for studying systems with reactivity such as split amorphous $\mathrm{SiO}_2$\cite{zou2012investigation}. Firstly, charge equilibrium of the amorphous silica block was carried out under the NPT ensemble, which differs from the previous one in that a reactive force field was used and the time step was 0.5 fs in \textit{real} unit. The next step is to apply deformation in the z-direction of the system so that the amorphous silica block gradually cracks, applying an engineering strain rate of 5 $\times 10^{-5}$/fs. After running for 22,000 steps in the NVT system at 300 K, we obtained a natural-cracked amorphous silica slit model, as shown in Fig.~\ref{fgr:model}b, with the cell dimension of $19.968 \times 20.128 \times 46.663$ \AA$^3$. Temperature and pressure are controlled with a Nos\'e-Hoover thermostat and barostat with relaxation times of 100 fs and 1000 fs, respectively.

\subsection{GCMC-MD simulations}
To investigate the adsorption characteristics of \ce{CO2} in amorphous silica under aqueous conditions, we set up a series of slit models with different hydrated conditions such as 1 wt\%, 2 wt\%, 5 wt\%, 10 wt\%, 15 wt\%. To create different water contents, we first performed the GCMC MD simulation on the obtained slit model to absorb different amounts of the water molecules. The empirical force filed of rigid four-site TIP4P/2005\cite{abascal2005general} was assigned to water. The Transferable Potential for Phase Equilibria (TraPPE)\cite{potoff2001vapor} force field was taken for modelling the \ce{CO2} molecules. The interactions between small molecules (\ce{H2O} and \ce{CO2}) and $\mathrm{SiO}_2$ was represented by combining the Vashishta and TIP4P/2005 and TraPPE force fields. The LJ parameters were obtained by using Lorentz-Berthelot mixing rule\cite{wisniak2010daniel} based on
\begin{equation}
\varepsilon_{i j}=\sqrt{\varepsilon_{i i} \times \varepsilon_{i j}},
\end{equation}
and
\begin{equation}
\sigma_{i j}=\frac{1}{2}\left(\sigma_{i i}+\sigma_{i j}\right).
\end{equation}
t
The parameters of the LJ potentials for \ce{SiO2}, \ce{H2O}, and \ce{CO2} are shown in Table~\ref{tbl:lj-potentials}. The cutoff for the long-range Coulomb and Van der Waals interactions was set as 10 \AA. 
\begin{table}[h]
\small
  \caption{\ Lennard-Jones potential parameters between $\mathrm{SiO}_2$, \ce{H2O}, and \ce{CO2} atoms in \textit{metal} unit. The $O_{Si}$, $O_{H}$, and $O_{C}$ are representing for the oxygen atoms belong to the $\mathrm{SiO}_2$, \ce{H2O}, and \ce{CO2} molecules, respectively.}
  \label{tbl:lj-potentials}
  \begin{tabular*}{0.48\textwidth}{@{\extracolsep{\fill}}lll}
    \hline
    Atom Pair&$\epsilon (eV)$&$\sigma ($\AA$)$ \\
    \hline
    Si-Si&0.0040&3.690\\
    $O_{Si}-O_{Si}$&0.0023&3.091\\
    $O_{H}-O_{H}$&0.0080 &3.159\\
    $O_{C}-O_{C}$&0.0179&2.626\\
    C-C&0.0106&2.811\\
    \hline
  \end{tabular*}
\end{table}

The simulation protocol of the sequential adsorption of \ce{H2O} and \ce{CO2} onto the amorphous $\mathrm{SiO}_2$ slit pores using a combination of Grand Canonical Monte Carlo (GCMC) methods and MD for temperature control. The process begins with the adsorption of water, followed by \ce{CO2} adsorption, with each stage carefully parameterized and executed. The simulation defines a specific region, slightly smaller than the initial simulation box (reduced by 0.1 \AA\ in all dimensions), to ensure that the interactions are confined within the area of interest, avoiding boundary effects. GCMC operations are performed every 100 steps, attempting 100 insertion or deletion operations per cycle, governed by a chemical potential of -0.5 eV with a fluctuation range of $\pm$0.1 eV. A temperature of 310 K is maintained during the GCMC operations, and the insertion attempts are scaled by a factor of $\frac{5}{3}$, effectively modifying the acceptance criteria for the insertion process. Additionally, molecular geometry corrections are applied to ensure structural consistency using the \textit{shake} command. Simultaneously, the dynamics of the water molecules are controlled using the NVT ensemble, maintaining the temperature at 310 K with a Nos\'e-Hoover thermostat at a coupling time of 0.1 ps in the \textit{metal} unit. The temperature computation for the \ce{H2O} group is dynamically updated to account for the changes in molecule numbers caused by GCMC operations. The $\mathrm{SiO}_2$ matrix is fixed and does not move, ensuring a stable background for the water molecules interactions. The time step is set to 1 fs for the NVT runs to achieve high accuracy in the dynamics calculations, and thermodynamic properties are recorded every 100 steps. The NVT simulation runs for 5000 steps, corresponding to 5 ps, to capture the short-term equilibrium behavior of the water molecules under the influence of chemical potential adjustments and dynamic temperature regulation. 

In the subsequent stage, \ce{CO2} adsorption is conducted using a similar GCMC framework, with \ce{CO2} molecules being introduced iteratively over 300 cycles. Each cycle consists of 500 simulation steps, allowing the system to equilibrate with the \ce{CO2} reservoir. Once the GCMC adsorption process is completed, the whole system is subjected to classical MD simulations of up to 1 ns under the NVT ensemble to obtain the diffusion characteristics of small molecules in the slit. The GCMC MD simulations were performed using the LAMMPS (Nov 21, 2023 version) package\cite{plimpton1995fast}.

\subsection{AIMD simulations}
Ab initio MD simulations were performed to study the interactions between small molecules {\ce{H2O} and \ce{CO2}} and amorphous $\mathrm{SiO}_2$ matrix. All calculations were conducted using the electronic structure module Quickstep in the CP2K version 9.1 software package \cite{kuhne2020cp2k}. The propagation of the classical nuclei was performed using \textit{ab initio} Born-Oppenheimer MD with a 0.5~fs timestep. At each step of the MD, the electronic orbitals were fine-tuned to the Born-Oppenheimer surface using an orbital transformation (OT) method \cite{vandevondele2003efficient} with a convergence criterion of $1 \times 10^{-7}$ a.u.%(= 1 Ha) 
The conjugate gradient minimizer was employed within the OT method to perform the optimization. %The \text{FULL_ALL} preconditioner was used with the minimization process. 
The Gaussian and plane waves method\cite{lippert1999gaussian} expanded the wave function in the Gaussian double-zeta valence polarized (DZVP) basis set \cite{vandevondele2007gaussian}. Auxiliary plane waves were employed to expand the electron density up to a cutoff of 400 Ry. One of the density functional approximations, Perdew-Burke-Ernzerhof (PBE) \cite{perdew1996generalized}, was used with Grimme dispersion corrections \cite{grimme2006semiempirical} denoted as D3 here. Adopting empirical dispersion correction to density functionals has shown superiority in optimizing water properties at ambient conditions\cite{knight2012multiscale}. Goedecker-Teter-Hutter pseudopotentials were utilized to effectively handle the core electrons \cite{goedecker1996separable}. The AIMD simulations were performed up to 15 ps in the NVT ensemble by using Nose'-Hoover (3 chains) thermostat at the time constant of 100 fs.
%The combined application of these methodologies is expected to yield a comprehensive framework for understanding the complex interplay of structural heterogeneity, surface chemistry, and external conditions that govern \ce{CO2} adsorption and diffusion in amorphous silica. By integrating the results from GCMC, MD, and AIMD simulations, this study aims to provide a holistic understanding of CO2 dynamics in heterogeneous porous media, thereby guiding the design of new silica-based materials with optimized properties for enhanced CO2 capture and storage. The outcomes of this research will contribute not only to the theoretical foundation of CO2 sequestration but also to the practical development of efficient storage strategies, ultimately advancing the broader field of CCUS and supporting global efforts to mitigate greenhouse gas emissions.
%\paragraph{This is the next level heading.~~} For this level please use \texttt{\textbackslash paragraph}. These headings should also end in a full point.
\subsection{Analysis details}
The excess adsorption of CO$_2$ ($n_\mathrm{ex}$) was calculated from MD trajectories by considering the difference between the absolute amount adsorbed and the amount that would be present in the same volume of bulk. For our system at supercritical conditions ($T = 310$ K, $P = 10$ MPa), the excess adsorption was determined using:

\begin{equation}
n_\mathrm{ex} = n_\mathrm{abs} - \frac{\rho_\mathrm{bulk} \times V_\mathrm{ads}}{m}
\end{equation}
where $n_\mathrm{abs}$ is the absolute adsorption (mmol/g), $\rho_\mathrm{bulk}$ is the bulk fluid density ($\mathrm{g} / \mathrm{cm}^3$), $V_\mathrm{ads}$ is the accessible pore volume ($\mathrm{cm}^3$), and $m$ is the mass of the adsorbent (g). The accessible pore volume was determined from the difference between the split and original simulation box volumes. The absolute adsorption was calculated
from the number of CO$_2$ molecules $\left(\mathrm{N}_{\mathrm{CO}_2}\right)$ in the simulation box:
\begin{equation}
n_\mathrm{abs} = \frac{N_\mathrm{CO_2} \times 1000}{m_\mathrm{ads} \times N_\mathrm{A}}
\end{equation}
where $N_\mathrm{A}$ is Avogadro number.

The bulk fluid density ($\rho_\mathrm{bulk}$) was calculated using the Peng-Robinson equation of state (PR-EOS)\cite{peng1976new}:

\begin{equation}
p = \frac{RT}{V_\mathrm{m} - b} - \frac{a\alpha}{V_\mathrm{m}^2 + 2bV_\mathrm{m} - b^2}
\end{equation}
where $p$ is the system pressure, $R$ is the universal gas constant (8.314 J mol$^{-1}$ K$^{-1}$), $T$ is the absolute temperature, and $V_\mathrm{m}$ represents the molar volume. The parameters $a$ and $b$ were determined from the critical properties of CO$_2$ ($T_\mathrm{c}$ = 304.13 K, $P_\mathrm{c}$ = 7.38 MPa).
%\begin{equation}
%a = 0.45724\frac{R^2T_\mathrm{c}^2}{P_\mathrm{c}}
%\end{equation}
%\begin{equation}
%b = 0.07780\frac{RT_\mathrm{c}}{P_\mathrm{c}}
%\end{equation}
%The temperature-dependent attractive term $\alpha$ was calculated as:
%\begin{equation}
%\alpha = [1 + \kappa(1-\sqrt{T/T_\mathrm{c}})]^2
%\end{equation}
%where $\kappa$ is correlated to the acentric factor ($\omega$ = 0.239 for CO$_2$):
%\begin{equation}
%\kappa = 0.37464 + 1.54226\omega - 0.26992\omega^2
%\end{equation}
The physical significance of these parameters lies in their representation of molecular interactions: the parameter $b$ accounts for the molecular volume exclusion effect, while the product $a\alpha$ characterizes the temperature-dependent attractive forces between molecules. This equation of state has been shown to provide reliable predictions for both vapor-liquid equilibria and volumetric properties of non-polar and slightly polar fluids, particularly in the critical and supercritical regions.

The self-diffusion coefficient ($D^s$) of CO$_2$ molecules was evaluated through MD simulations in the canonical ($NVT$) ensemble at 310 K and 100 bar. The mean square displacement (MSD) was calculated using Einstein relation:

\begin{equation}
D^s = \frac{1}{2dN}\lim_{t\to\infty}\frac{d}{dt}\left\langle\sum_{i=1}^N (r_i^{\mathrm{CO_2}}(t) - r_i^{\mathrm{CO_2}}(0))^2\right\rangle
\end{equation}
where $d$ is the system spacial dimension, $N$ is the number of CO$_2$ molecules, and $r_i^{\mathrm{CO_2}}(t)$ represents the position of the $i$th CO$_2$ molecule at time $t$. The self-diffusion coefficient was obtained from the slope of the MSD versus time in the linear regime, where the system exhibited normal diffusive behaviour.
\section{Results and discussion}
\subsection{CO$_2$ adsorption kinetics}
The excess adsorption curves of CO$_2$ in hydrated amorphous silica slit pores exhibit several distinctive features that provide valuable insights into the adsorption mechanism. Most notably, a clear water content-dependent behaviour is observed, where the equilibrium excess adsorption capacity systematically decreases with increasing water content, from approximately 7.6 mmol/g at 1 wt\% to 2.6 mmol/g at 15 wt\%. This significant reduction (approximately 66 \%) in CO$_2$ uptake capacity suggests a strong competitive effect between \ce{H2O} and \ce{CO2} for available adsorption sites. The adsorption of \ce{CO2} in the amorphous silica slit model at low water content is consistent with the data on \ce{CO2} adsorption in an activated carbon slit model measured experimentally and by GCMC in the work of Ref.~\cite{gomaa2022experimental}, which is roughly on the order of 8 mmol/g. The slight decrease in our adsorption amount is the effect of pre-adsorbed water molecules.
\begin{figure}[h]
\centering
  \includegraphics[height=6.0cm]{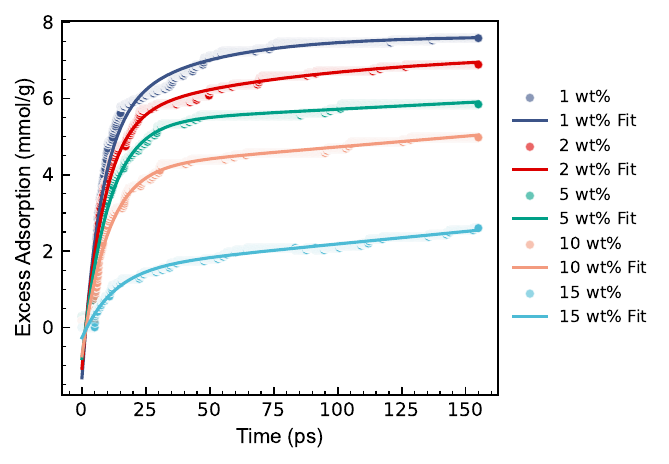}
  \caption{The \ce{CO2} adsorption in amorphous silica slit model at different water contents at 310 K and 10 MPa. double exponential model was used to fit the adsorption curves.}
  \label{fgr:adsorption}
\end{figure}

The adsorption kinetics can be characterized by two distinct regimes across all water contents. The initial stage (0-25 ps) shows a steep uptake rate, indicating rapid \ce{CO2} adsorption on readily accessible surface sites. This is followed by a gradual approach to equilibrium (25-150 ps), where the adsorption rate significantly decreases. Interestingly, the time required to reach equilibrium (approximately 100 ps) remains relatively consistent across different water contents, despite the substantial differences in equilibrium capacity. A particularly noteworthy feature is the evolution of the curve shapes with increasing water content. At low water contents (1-2 wt\%), the curves exhibit a sharp initial rise followed by a smooth transition to equilibrium. However, as water content increases, the initial uptake becomes less steep, and the transition to equilibrium becomes more gradual, especially evident in the 10 wt\% and 15 wt\% cases. This behaviour suggests that higher water content not only reduces the available adsorption sites but also modifies the adsorption kinetics by creating additional diffusion barriers for \ce{CO2} molecules.

The data points show some fluctuations around the fitted curves, particularly noticeable at higher water contents, which can be attributed to the dynamic nature of water-\ce{CO2} interactions and the continuous reorganization of water molecules within the confined space. These fluctuations provide evidence of the complex interplay between water, \ce{CO2}, and the silica surface in the nanoporous environment. These mechanisms were revealed later in the simulation results of AIMD.

The selection of an appropriate kinetic model is crucial for understanding the CO$_2$ adsorption mechanism in hydrated amorphous silica slit model. Initially, several conventional adsorption kinetic models were considered and evaluated: 1. Pseudo-first-order model (PFO)\cite{csahin2022adsorption}:
\begin{equation}
q_t=q_e\left(1-\exp ^{-k_1, t}\right)
\end{equation}
where $q_t$ and $q_e$ represent the amount of CO$_2$ adsorbed at time $t$ and at equilibrium, respectively, and $k_1$ is the first-order rate constant. 2. Pseudo-second-order model (PSO)\cite{narasimharao2022fe3o4}:
\begin{equation}
q_t=\frac{\left(q_e^2 \cdot k_2 \cdot t\right)}{1+q_e \cdot k_2 \cdot t}
\end{equation}
where $k_2$ is the second-order rate constant. 3. Avrami kinetic model\cite{shafeeyan2015modeling}:
\begin{equation}
q_t=q_e\left(1-\exp ^{-k_{a v} \cdot t^{n_{a v}}}\right)
\end{equation}
where $k_{av}$ is the kinetic rate constant and $n_{av}$ is the Avrami exponent representing the dimensionality of growt\%h.

However, careful examination of the adsorption curves revealed that these conventional models were inadequate to capture the complex nature of CO$_2$ adsorption in our hydrated system. The presence of water molecules introduces additional complexity to the adsorption process, as evidenced by the distinct fast and slow uptake regions observed in our data. This observation led us to adopt the double exponential (DE) model\cite{marczewski2010application}:
\begin{equation}
q_t=q_e\left[A_1\left(1-\exp ^{-k_1 \cdot t}\right)+A_2\left(1-\exp ^{-k_2 \cdot t}\right)\right]
\end{equation}
where $k_1$ and $k_2$ are the rate constants for fast and slow adsorption processes, $A_1$ and $A_2$ are the relative contributions of each process (A1 + A2 = 1). The DE model was selected based on several key considerations. The model accounts for two distinct adsorption processes: a rapid initial uptake ($k_1$) representing direct surface adsorption and a slower secondary process ($k_2$) reflecting diffusion-limited adsorption, as shown in Fig.~\ref{fgr:adsorption}. 

The model demonstrated excellent agreement with the GCMC simulation results across all water contents, yielding high $R^2$ values approximately above 0.98 and demonstrating its robustness in describing the adsorption process across different hydration conditions. The quality of the fitting increased slightly with increasing water content.
\begin{figure}[h]
\centering
  \includegraphics[height=13cm]{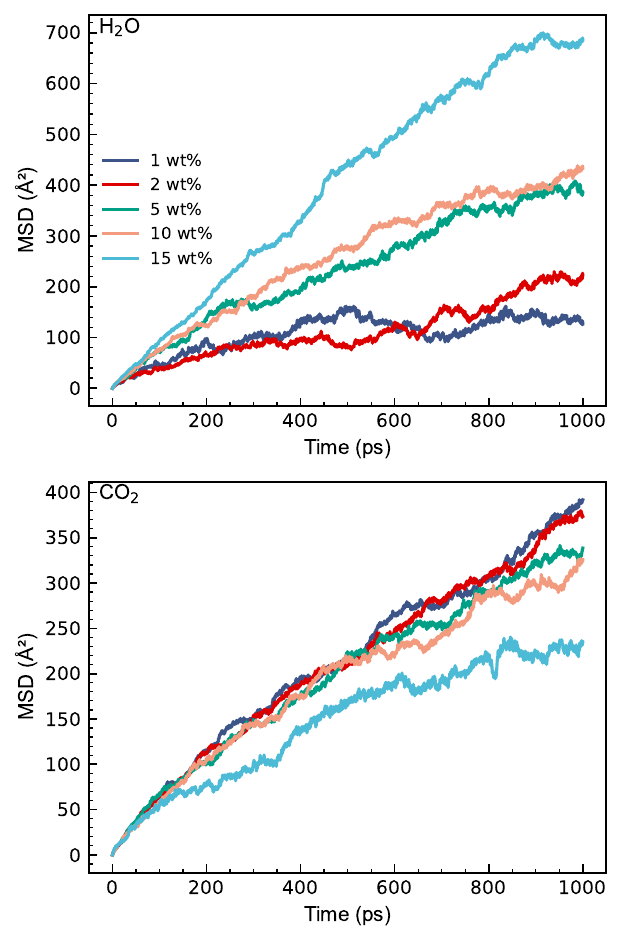}
  \caption{Bulk summation of the mean square displacement of \ce{H2O} (Top panel) and \ce{CO2} (Bottom panel) molecules in amorphous silica slit model at different water contents at 310 K and 10 MPa. }
  \label{fgr:diffusion}
\end{figure}

\subsection{Diffusion properties of \ce{H2O} and \ce{CO2}}
The mean square displacement (MSD) analysis reveals distinct diffusion behaviours of \ce{H2O} and \ce{CO2} molecules under varying water contents in the amorphous silica slit model. These classical MD results provide initial insights into the complex molecular transport mechanisms, while setting the stage for more detailed quantum mechanical investigations below through AIMD simulations. 

For \ce{H2O} molecules, a pronounced enhancement in mobility is observed with increasing water content from 1 wt\% to 15 wt\%, as evidenced by the significant increase in MSD values from approximately 150 Å² to 700 Å² at 1000 ps. At low water contents (1-2 wt\%), the MSD curves tend to plateau after 400-600 ps, suggesting restricted molecular motion. This behaviour can be attributed to the strong interactions between water molecules and the Si-O-Si network of the amorphous silica framework, where water molecules form hydrogen bonds with the oxygen atoms in the silica structure, effectively constraining their diffusion. Although classical MD captures this general diffusion trend, AIMD simulations would be particularly valuable in revealing the precise nature of these hydrogen bonds and their impact on electron density distributions at the water-silica interface.

In contrast, at higher water contents (10-15 wt\%), the continuously rising MSD curves with steeper slopes indicate enhanced molecular mobility. This enhancement likely results from the formation of continuous water networks due to the large amount of water molecules, where water-water interactions become dominant over water-silica interactions, facilitating water transport through the porous structure. The transition from surface-dominated to network-dominated transport suggests complex changes in the electronic structure of the system, which can be further elucidated through AIMD simulations to understand the quantum mechanical aspects of hydrogen bond network dynamics.

Intriguingly, \ce{CO2} molecules exhibit an inverse relationship with water content. The diffusion of \ce{CO2} is progressively hindered as water content increases, demonstrated by the decrease in MSD values from 380 Å² at 1 wt\% to 230 Å² at 15 wt\% after 1000 ps. Notably, the MSD values of \ce{CO2} are consistently lower than those of \ce{H2O} under all conditions, indicating more restricted diffusion of \ce{CO2} molecules. At higher water contents (10-15 wt\%), the \ce{CO2} MSD curves show a tendency to plateau beyond 600 ps, suggesting that the formed water networks may create additional barriers for \ce{CO2} transport.

These findings reveal a dual-regulation mechanism of water content on molecular transport in amorphous silica. While increased water content promotes water diffusion by transitioning from surface-dominated to network-dominated transport, it simultaneously suppresses \ce{CO2} mobility through spatial competition and pathway obstruction. This inverse regulation effect demonstrates the critical role of water content in controlling selective molecular transport through silica-based materials. Furthermore, this classical MD study provides a foundation for subsequent AIMD investigations, which can offer deeper insights into the quantum mechanical nature of molecular interactions, charge transfer processes, and the formation/breaking of chemical bonds that may occur during molecular transport.
\begin{figure}[h]
\centering
  \includegraphics[height=13cm]{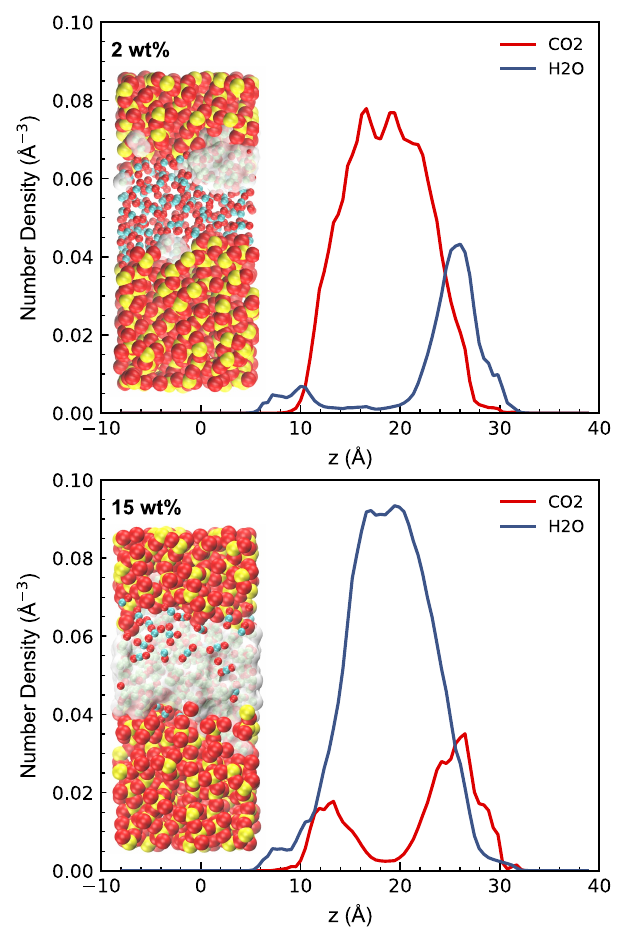}
  \caption{Number density profiles of \ce{H2O} and \ce{CO2} molecules in slit pores at 2 wt\% and 15 wt\% water contents. Silicon atoms are coloured in yellow, oxygen atoms different molecules are red, hydrogen atoms are white, and carbon atoms are cyan. The screenshots show the adsorption occurrences of \ce{H2O} and \ce{CO2} molecules of the final frames from the 1~ns classical MD simulations.}
  \label{fgr:sites}
\end{figure}

\subsection{Adsorption sites of \ce{H2O} and \ce{CO2}}
The density profiles reveal distinct distribution patterns of \ce{H2O} and \ce{CO2} molecules within the nanopore under different water contents. Representative results for one low (2 wt\%) and one high (15 wt\%) moisture content are selected for analysis here. At 2~wt\% water content, \ce{CO2} molecules predominantly occupy the central region of the pore ($z \approx$ 15-25 \AA) with a high density peak ($\approx$ 0.08~\AA$^{-3}$), while water molecules are mainly localized near the pore surface ($z \approx$ 25-30~\AA) with relatively lower density ($\approx$ 0.04~\AA$^{-3}$). This initial distribution reflects the strong hydrophilic interactions between water molecules and silicon-oxygen groups through hydrogen bonding, while \ce{CO2} molecules occupy the central region to maximize their configurational entropy. At this pore scale, the distribution of \ce{CO2} molecules also exhibits a bimodal structure similar to that exhibited by \ce{CH4} in our previous work\cite{shi2019competitive}, as shown in Fig.~\ref{fgr:sites}.

As the water content increases to 15 wt\%, a significant redistribution occurs through complex molecular interactions, as shown in Fig.~\ref{fgr:sites}. The water density in the central region increases substantially ($\approx$ 0.09~\AA$^{-3}$), forming a dominant water phase through extensive hydrogen bonding networks. Notably, \ce{CO2} molecules exhibit a unique dual-peak distribution: one near the one side of the surface ($z \approx$ 10-15 \AA) and another near the other side of the reigon ($z \approx$ 25-30 \AA). This redistribution mechanism can be attributed to several factors: First, the large number of the formation of ordered water structures in the central region creates energetically favourable interfacial sites near the surface for \ce{CO2} molecules, where they can interact with both partially exposed silicon atoms and the ordered water layer through their quadrupole moment. Second, the ordered water structure near the surface may create local hydrophobic pockets that can accommodate \ce{CO2} molecules while minimizing disruption to the water hydrogen bonding network. Furthermore, the nanoscale confinement enhances these surface-mediated interactions, making the surface region an energetically competitive location for \ce{CO2} molecules.

These observations demonstrate the competition between different molecular interactions - namely \ce{H2O}-surface hydrogen bonding, \ce{H2O}-\ce{H2O} hydrogen bonding networks, and \ce{CO2}-surface/\ce{CO2}-water interactions - governs the spatial organization of molecules within the nanopore. The observed behaviour suggests that at higher water contents, while the formation of structured water networks dominates the central pore region, it simultaneously creates unique interfacial environments that influence \ce{CO2} distribution. This understanding of competitive adsorption and molecular reorganization is crucial for optimizing \ce{CO2} capture and storage in nanoporous materials under realistic conditions where water is inevitably present.
\begin{figure}[h]
\centering
  \includegraphics[height=10cm]{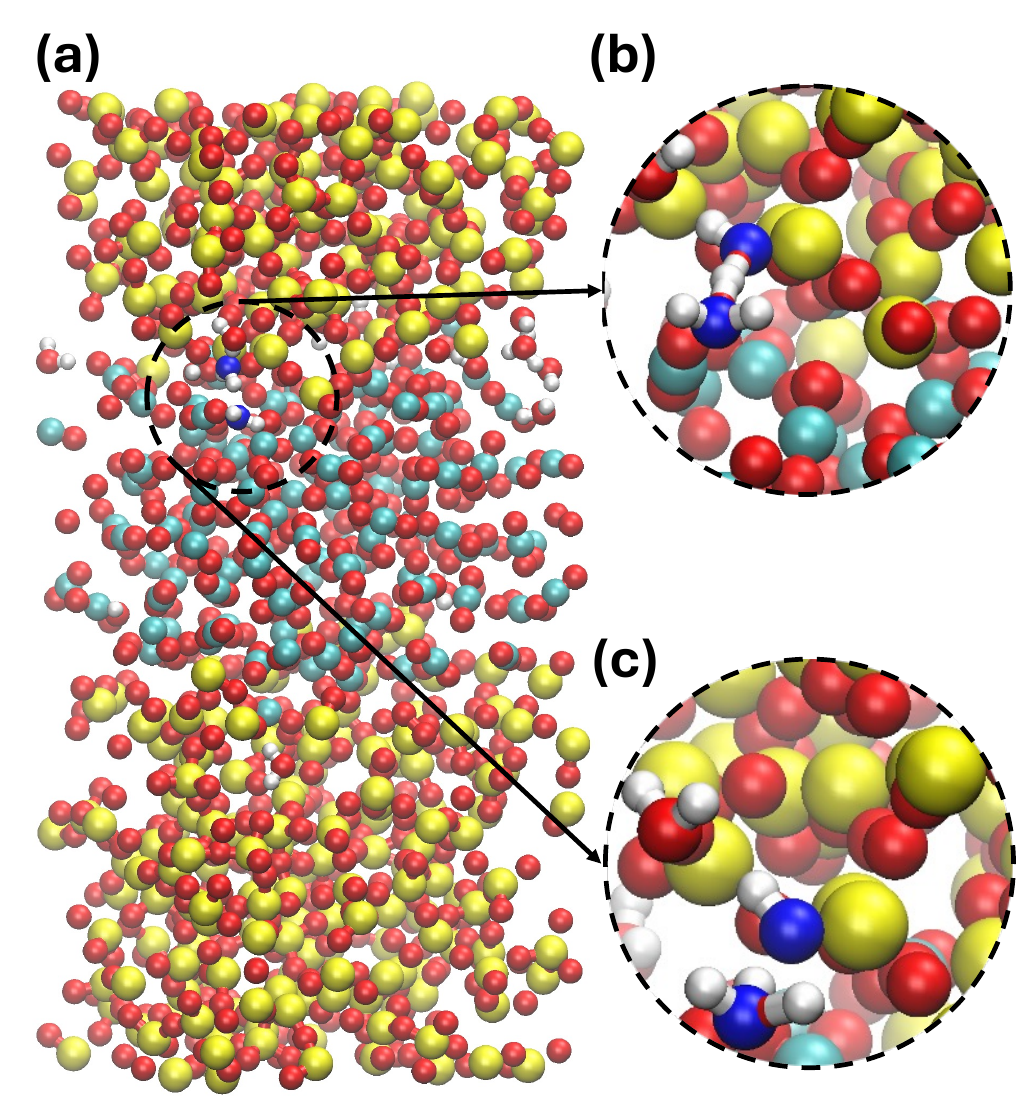}
  \caption{The formation of surface silanol groups and the frequently proton transfer of the representative 2 wt\% system during AIMD simulation. Silicon atoms are coloured in yellow, oxygen atoms different molecules are red, hydrogen atoms are white, and carbon atoms are cyan. The transit structure of \ce{OH-}-\ce{H3O+} complex (b) and the formation of the \ce{H3O+} ion. The screenshots were captured from the upto 15 ps AIMD simulations.}
  \label{fgr:proton}
\end{figure}

\subsection{Insights into the intermolecular interactions at the DFT scale}
Our AIMD simulations revealed a proton transfer mechanism mediated by water molecules in the hydrated silica nanopores during \ce{CO2} adsorption. The trajectory analysis showed the formation of a transient \ce{OH-}-\ce{H3O+} complex through the coordination of two water molecules, as shown in Fig.~\ref{fgr:proton} (b). This complex exhibited notable structural features with an O-O distance of 2.51 \AA, characteristic of a strong hydrogen bonding interaction. This structural feature aligns well with previous theoretical studies \cite{izvekov2005ab} reporting O-O distances of 2.4-2.5 \AA\ for proton-sharing water complexes, confirming the formation of a stable \ce{OH-}-\ce{H3O+} species instead of the tranditional Zundel cation\cite{zundel1969hydration}. The subsequent proton transfer process proceeded via a Grotthuss-like mechanism\cite{agmon1995grotthuss}, where one water molecule donated a proton to form \ce{H3O+}, as shown in Fig.~\ref{fgr:proton} c. While the remaining \ce{OH-} simultaneously coordinated with a surface silicon atom of the amorphous silica framework. 
\begin{figure}[h]
\centering
  \includegraphics[height=10cm]{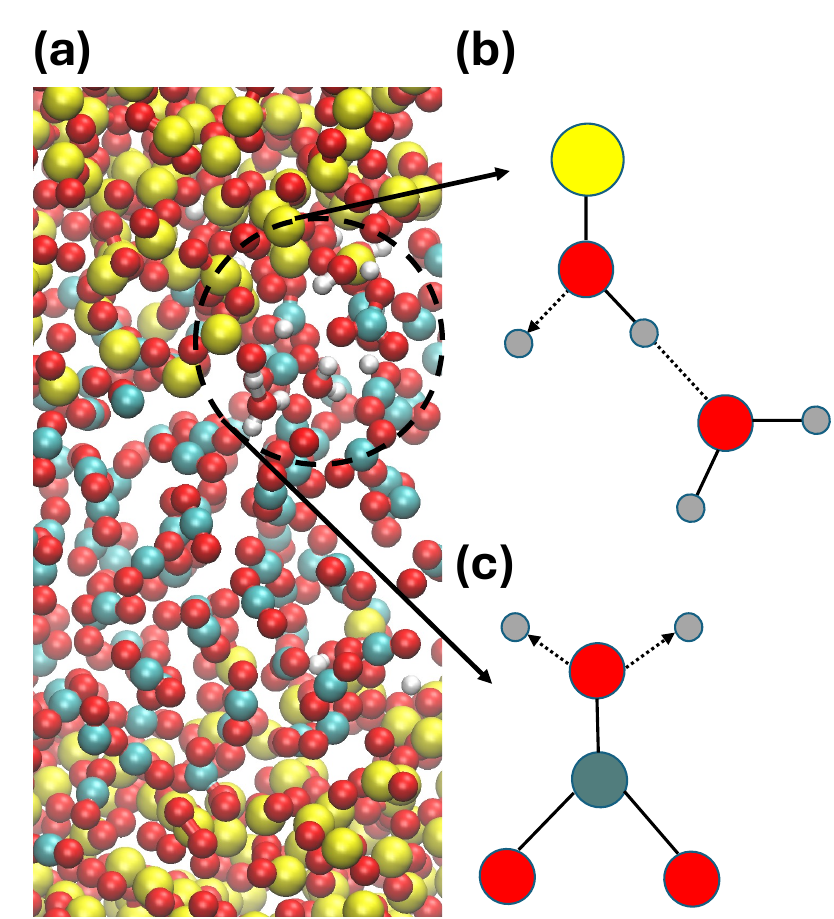}
  \caption{Hydrogen bonding networks and \ce{CO2} hydration processes of the 2 wt\% system during AIMD simulations. The sketches of the process of silicon silanol group on the surface linking a water molecule via hydrogen bonding (b), as well as the eventual formation of \ce{CO3^{2-}} ion (c). Silicon atoms are coloured in yellow, oxygen atoms different molecules are red, hydrogen atoms are white, and carbon atoms are cyan. The transit structure of \ce{OH-}-\ce{H3O+} complex (b) and the formation of the \ce{H3O+} ion. The screenshots were captured from the upto 15 ps AIMD simulations.}
  \label{fgr:co2-hydrated}
\end{figure}

The driving force for later proton transfer event can be attributed to the strong Lewis acidity of framework Si atoms in the silica structure. When two water molecules come into close proximity (O-O distance of 2.4-2.5 \AA), the electron-deficient Si atom from an existing Si-O framework bond exhibits a strong attraction towards the hydroxyl group of one water molecule. This Si···\ce{OH-} interaction creates an electronic driving force that facilitates the separation of \ce{OH-} and \ce{H+} from the water molecule. The stabilization of the hydroxyl group by the framework Si atom effectively lowers the energy barrier for proton transfer, enabling the formation of the transit \ce{OH-}-\ce{H3O+} complex and subsequent proton hopping to the neighboring water molecule. From an electronic structure perspective, the undercoordinated Si centers (dangling sites) on the amorphous silica surface act as strong electron density acceptors. The interaction between such an electron-deficient dangling Si site and the electron-rich oxygen of the \ce{OH-} creates a stabilizing effect through orbital interactions. This surface Si···\ce{OH-} interaction effectively anchors the hydroxyl group, providing a thermodynamically favorable pathway for water dissociation and subsequent proton transfer

The formation of surface silanol groups through water dissociation fundamentally alters the local chemical environment of the amorphous silica surface. These newly formed Si-\ce{OH-} groups modify the surface electronic structure and local charge distribution, potentially enhancing the binding strength with \ce{CO2} molecules; second, they act as hydrogen bond donors/acceptors, creating an extended hydrogen bonding network that can stabilize adsorbed \ce{CO2} molecules. The proton transfer dynamics we observed suggest that these surface \ce{OH-} are not static entities but rather participate in dynamic proton exchange, which could facilitate \ce{CO2} capture through cooperative binding mechanisms. This surface hydroxylation effect transforms the originally less reactive silica surface into a more chemically active interface for \ce{CO2} adsorption, where both electrostatic interactions and hydrogen bonding contribute to the enhanced adsorption capacity, which is consistent with the experimental results by \citeauthor{gomaa2023experimental}\cite{gomaa2023experimental}.

After the formation of silanol groups on the surface, we observed the formation of hydrogen bonds with the surrounding water molecules as shown in Fig.~\ref{fgr:co2-hydrated} b. The formation of hydrogen bonds between surface silanol groups and water molecules significantly influences the spatial organization of confined species. Our simulations reveal that these silanol-water hydrogen bonds induce the formation of a structured water layer adjacent to the silica surface, as evidenced by the density distribution analysis from Fig.~\ref{fgr:sites}. This ordered hydration layer modifies the local chemical environment near the surface, affecting the spatial distribution and accessibility of \ce{CO2} molecules within the confined space. In addition, the spontaneous hydration of \ce{CO2} was also captured by AIMD simulations. The trajectory analysis revealed that the \ce{CO2} hydration process proceeds via a concerted mechanism involving multiple water molecules. Specifically, we observed the formation of carbonic acid (\ce{H2CO3}) through nucleophilic attack of a water molecule on the carbon center of \ce{CO2}, followed by proton transfer events that ultimately led to stable carbonate (\ce{CO3$^{2-}$}) species, as shown in Fig.~\ref{fgr:co2-hydrated} c. The AIMD simulations capture the step-by-step mechanism of carbonic acid formation, starting with the hydration of \ce{CO2} to form carbonic acid, followed by its dissociation into bicarbonate (\ce{HCO3-}) and carbonate (\ce{CO3^{2-}}) ions. The O-H bond breaking and formation during this process was facilitated by the hydrogen bonding network in the confined water layer, which effectively lowered the activation barrier for \ce{CO2} hydration.

This \ce{CO2} hydration process represents an additional pathway in the overall adsorption mechanism. The formation of carbonate species provides an alternative state for \ce{CO2} in the confined space, where these hydrated species can interact with both the surrounding water molecules and surface silanol groups through hydrogen bonding networks. This finding suggests that the presence of confined water in silica nanopores enables \ce{CO2} hydration, potentially diversifying the \ce{CO2} capture mechanisms in the system.

In summary, our AIMD simulations reveal that \ce{CO2} molecules confined in hydrated silica nanopores can undergo spontaneous hydration, leading to the formation of carbonate species. This finding extends beyond the conventional understanding of purely physical adsorption mechanisms, demonstrating the dynamic interplay between \ce{CO2}, confined water, and the silica surface at the molecular level. The observation of formation of the silanol groups, proton tranfer, the \ce{CO2} hydration in nanoporous environments provides atomic-scale insights into the complex chemical processes occurring in practical carbon capture systems, highlighting the dual role of water as both a reaction medium and an active participant in \ce{CO2} capture.

\section*{Conclusions}
This study provides molecular insights into the complex interplay between \ce{CO2}, \ce{H2O}, and amorphous silica surfaces in slit nanopores. Our multi-scale computational approach reveals several key findings. The confined \ce{H2O} molecules significantly impacts \ce{CO2} uptake capacity, with adsorption kinetics following a double exponential model that reflects distinct fast and slow uptake processes. Besides, molecular diffusion analysis demonstrates an inverse relationship between \ce{H2O} and \ce{CO2} mobility, where increased water content enhances water diffusion while hindering \ce{CO2} transport. Furthermore, density distribution analysis reveals a water-content-dependent reorganization of adsorption sites, with \ce{CO2} transitioning from central pore regions to surface-proximate locations at higher water contents.

Crucially, our AIMD simulations uncover previously unrecognized chemical processes in the confined space. We observe proton transfer mechanisms leading to surface silanol formation, with characteristic O-O distances of 2.4-2.5 \AA\ in the transient two water bonding complex. The simulations also capture spontaneous \ce{CO2} hydration events, resulting in \ce{CO3^{2-}} species formation through water-mediated reaction pathways. These findings demonstrate that confined water plays dual roles: as a spatial competitor for adsorption sites and as a reaction medium enabling chemical transformations of \ce{CO2}.

These insights extend beyond conventional physisorption models and provide a quantum mechanical level foundation for understanding \ce{CO2} capture in hydrated nanoporous materials. The identified mechanisms of surface modification and \ce{CO2} transformation suggest new strategies for optimizing carbon capture technologies, particularly in realistic conditions where water is invariably present.

\section*{Acknowledgements}
We are grateful for the computational resources funded by the National Natural Science Foundation of China (No. 52006243and No. 51936001), the Shandong Excellent Young Scientist (Overseas) Program (No. 2024HWYQ-050), the Qingdao Postdoc Research Program (No. QDBSH20240201031), the Natural Science Foundation of Shandong Province (No. ZR2020QE197), and the Fundamental Research Funds for the Central Universities (No. 22CX06027A).

\section*{Author contributions}
JS, TZ, CZ, SS and LG participated in the research design. JS performed the molecular simulations, analyzed the data and wrote the original manuscript. TZ, CZ, SS, and LG revised the manuscript. TZ, LG provided computational resources. SS and LG conceived the project, supervised the research. All authors contributed to the discussion of the results and the revision of the manuscript and approved the submitted version.

\section*{Conflicts of interest}
There are no conflicts of interest to declare.

\section*{Data availability}
The data supporting this article is openly available from the research data repository, Harvard Dataverse, at xxx. (The dataset will be deposited later.)

%%%END OF MAIN TEXT%%%

\balance

\renewcommand\refname{References}

%%%REFERENCES%%%
\bibliography{references} %You need to replace "rsc" on this line with the name of your .bib file

\providecommand*{\mcitethebibliography}{\thebibliography}
\csname @ifundefined\endcsname{endmcitethebibliography}
{\let\endmcitethebibliography\endthebibliography}{}
\begin{mcitethebibliography}{64}
\providecommand*{\natexlab}[1]{#1}
\providecommand*{\mciteSetBstSublistMode}[1]{}
\providecommand*{\mciteSetBstMaxWidthForm}[2]{}
\providecommand*{\mciteBstWouldAddEndPuncttrue}
  {\def\EndOfBibitem{\unskip.}}
\providecommand*{\mciteBstWouldAddEndPunctfalse}
  {\let\EndOfBibitem\relax}
\providecommand*{\mciteSetBstMidEndSepPunct}[3]{}
\providecommand*{\mciteSetBstSublistLabelBeginEnd}[3]{}
\providecommand*{\EndOfBibitem}{}
\mciteSetBstSublistMode{f}
\mciteSetBstMaxWidthForm{subitem}
{(\emph{\alph{mcitesubitemcount}})}
\mciteSetBstSublistLabelBeginEnd{\mcitemaxwidthsubitemform\space}
{\relax}{\relax}

\bibitem[Bouwer(2013)]{bouwer2013projections}
L.~M. Bouwer, \emph{Risk Analysis}, 2013, \textbf{33}, 915--930\relax
\mciteBstWouldAddEndPuncttrue
\mciteSetBstMidEndSepPunct{\mcitedefaultmidpunct}
{\mcitedefaultendpunct}{\mcitedefaultseppunct}\relax
\EndOfBibitem
\bibitem[Masson-Delmotte \emph{et~al.}(2021)Masson-Delmotte, Zhai, Pirani, Connors, P{\'e}an, Berger, Caud, Chen, Goldfarb, Gomis,\emph{et~al.}]{masson2021climate}
V.~Masson-Delmotte, P.~Zhai, A.~Pirani, S.~L. Connors, C.~P{\'e}an, S.~Berger, N.~Caud, Y.~Chen, L.~Goldfarb, M.~I. Gomis \emph{et~al.}, \emph{Contribution of working group I to the sixth assessment report of the intergovernmental panel on climate change}, 2021, \textbf{2}, 2391\relax
\mciteBstWouldAddEndPuncttrue
\mciteSetBstMidEndSepPunct{\mcitedefaultmidpunct}
{\mcitedefaultendpunct}{\mcitedefaultseppunct}\relax
\EndOfBibitem
\bibitem[Boubaker \emph{et~al.}(2024)Boubaker, Liu, Mu, and Zhan]{boubaker2024carbon}
S.~Boubaker, Z.~Liu, Y.~Mu and Y.~Zhan, \emph{Risk Analysis}, 2024\relax
\mciteBstWouldAddEndPuncttrue
\mciteSetBstMidEndSepPunct{\mcitedefaultmidpunct}
{\mcitedefaultendpunct}{\mcitedefaultseppunct}\relax
\EndOfBibitem
\bibitem[Gabrielli \emph{et~al.}(2020)Gabrielli, Gazzani, and Mazzotti]{gabrielli2020role}
P.~Gabrielli, M.~Gazzani and M.~Mazzotti, \emph{Industrial \& Engineering Chemistry Research}, 2020, \textbf{59}, 7033--7045\relax
\mciteBstWouldAddEndPuncttrue
\mciteSetBstMidEndSepPunct{\mcitedefaultmidpunct}
{\mcitedefaultendpunct}{\mcitedefaultseppunct}\relax
\EndOfBibitem
\bibitem[Lal(2008)]{lal2008carbon}
R.~Lal, \emph{Philosophical Transactions of the Royal Society B: Biological Sciences}, 2008, \textbf{363}, 815--830\relax
\mciteBstWouldAddEndPuncttrue
\mciteSetBstMidEndSepPunct{\mcitedefaultmidpunct}
{\mcitedefaultendpunct}{\mcitedefaultseppunct}\relax
\EndOfBibitem
\bibitem[Gbadamosi \emph{et~al.}(2019)Gbadamosi, Junin, Manan, Agi, and Yusuff]{gbadamosi2019overview}
A.~O. Gbadamosi, R.~Junin, M.~A. Manan, A.~Agi and A.~S. Yusuff, \emph{International Nano Letters}, 2019, \textbf{9}, 171--202\relax
\mciteBstWouldAddEndPuncttrue
\mciteSetBstMidEndSepPunct{\mcitedefaultmidpunct}
{\mcitedefaultendpunct}{\mcitedefaultseppunct}\relax
\EndOfBibitem
\bibitem[Xie \emph{et~al.}(2024)Xie, Zhang, Zhao, Cao, Li, and Zhang]{10.1007/978-981-97-0268-8_34}
Z.-h. Xie, L.-h. Zhang, Y.-l. Zhao, C.~Cao, L.-x. Li and D.-p. Zhang, Proceedings of the International Field Exploration and Development Conference 2023, Singapore, 2024, pp. 434--446\relax
\mciteBstWouldAddEndPuncttrue
\mciteSetBstMidEndSepPunct{\mcitedefaultmidpunct}
{\mcitedefaultendpunct}{\mcitedefaultseppunct}\relax
\EndOfBibitem
\bibitem[Yang \emph{et~al.}(2010)Yang, Bai, Tang, Shari, and David]{yang2010characteristics}
F.~Yang, B.~Bai, D.~Tang, D.-N. Shari and W.~David, \emph{Petroleum Science}, 2010, \textbf{7}, 83--92\relax
\mciteBstWouldAddEndPuncttrue
\mciteSetBstMidEndSepPunct{\mcitedefaultmidpunct}
{\mcitedefaultendpunct}{\mcitedefaultseppunct}\relax
\EndOfBibitem
\bibitem[Talapatra(2020)]{talapatra2020study}
A.~Talapatra, \emph{Journal of Petroleum Exploration and Production Technology}, 2020, \textbf{10}, 1965--1981\relax
\mciteBstWouldAddEndPuncttrue
\mciteSetBstMidEndSepPunct{\mcitedefaultmidpunct}
{\mcitedefaultendpunct}{\mcitedefaultseppunct}\relax
\EndOfBibitem
\bibitem[Fentaw \emph{et~al.}(2024)Fentaw, Emadi, Hussain, Fernandez, and Thiyagarajan]{en17195000}
J.~W. Fentaw, H.~Emadi, A.~Hussain, D.~M. Fernandez and S.~R. Thiyagarajan, \emph{Energies}, 2024, \textbf{17}, 5000\relax
\mciteBstWouldAddEndPuncttrue
\mciteSetBstMidEndSepPunct{\mcitedefaultmidpunct}
{\mcitedefaultendpunct}{\mcitedefaultseppunct}\relax
\EndOfBibitem
\bibitem[Akhter \emph{et~al.}(2021)Akhter, Soomro, and Inglezakis]{akhter2021silica}
F.~Akhter, S.~A. Soomro and V.~J. Inglezakis, \emph{Journal of Porous Materials}, 2021, \textbf{28}, 1387--1400\relax
\mciteBstWouldAddEndPuncttrue
\mciteSetBstMidEndSepPunct{\mcitedefaultmidpunct}
{\mcitedefaultendpunct}{\mcitedefaultseppunct}\relax
\EndOfBibitem
\bibitem[Pettinari and Tombesi(2020)]{pettinari2020metal}
C.~Pettinari and A.~Tombesi, \emph{MRS Energy \& Sustainability}, 2020, \textbf{7}, E35\relax
\mciteBstWouldAddEndPuncttrue
\mciteSetBstMidEndSepPunct{\mcitedefaultmidpunct}
{\mcitedefaultendpunct}{\mcitedefaultseppunct}\relax
\EndOfBibitem
\bibitem[Navik \emph{et~al.}(2024)Navik, Wang, Ding, Qiu, and Li]{navik2024atmospheric}
R.~Navik, E.~Wang, X.~Ding, K.~Qiu and J.~Li, \emph{Environmental Chemistry Letters}, 2024,  1--40\relax
\mciteBstWouldAddEndPuncttrue
\mciteSetBstMidEndSepPunct{\mcitedefaultmidpunct}
{\mcitedefaultendpunct}{\mcitedefaultseppunct}\relax
\EndOfBibitem
\bibitem[Ramasamy \emph{et~al.}(2024)Ramasamy, Raj, Akula, and Nagarasampatti~Palani]{ramasamy2024leveraging}
N.~Ramasamy, A.~J. L.~P. Raj, V.~V. Akula and K.~Nagarasampatti~Palani, \emph{Environmental Science and Pollution Research}, 2024,  1--30\relax
\mciteBstWouldAddEndPuncttrue
\mciteSetBstMidEndSepPunct{\mcitedefaultmidpunct}
{\mcitedefaultendpunct}{\mcitedefaultseppunct}\relax
\EndOfBibitem
\bibitem[Chittick \emph{et~al.}(1969)Chittick, Alexander, and Sterling]{chittick1969preparation}
R.~Chittick, J.~Alexander and H.~Sterling, \emph{Journal of the Electrochemical Society}, 1969, \textbf{116}, 77\relax
\mciteBstWouldAddEndPuncttrue
\mciteSetBstMidEndSepPunct{\mcitedefaultmidpunct}
{\mcitedefaultendpunct}{\mcitedefaultseppunct}\relax
\EndOfBibitem
\bibitem[Dias and Alves(2022)]{dias2022silica}
L.~S. Dias and A.~K. Alves, \emph{Technological Applications of Nanomaterials}, 2022,  89--106\relax
\mciteBstWouldAddEndPuncttrue
\mciteSetBstMidEndSepPunct{\mcitedefaultmidpunct}
{\mcitedefaultendpunct}{\mcitedefaultseppunct}\relax
\EndOfBibitem
\bibitem[Adams(1975)]{doi:10.1080/00268977500100221}
D.~Adams, \emph{Molecular Physics}, 1975, \textbf{29}, 307--311\relax
\mciteBstWouldAddEndPuncttrue
\mciteSetBstMidEndSepPunct{\mcitedefaultmidpunct}
{\mcitedefaultendpunct}{\mcitedefaultseppunct}\relax
\EndOfBibitem
\bibitem[Khoshraftar \emph{et~al.}(2021)Khoshraftar, Ghaemi, and Mashhadimoslem]{khoshraftar2021evaluation}
Z.~Khoshraftar, A.~Ghaemi and H.~Mashhadimoslem, \emph{Iranian Journal of Chemical Engineering (IJChE)}, 2021, \textbf{18}, 64--80\relax
\mciteBstWouldAddEndPuncttrue
\mciteSetBstMidEndSepPunct{\mcitedefaultmidpunct}
{\mcitedefaultendpunct}{\mcitedefaultseppunct}\relax
\EndOfBibitem
\bibitem[Kn{\"o}fel \emph{et~al.}(2009)Kn{\"o}fel, Martin, Hornebecq, and Llewellyn]{doi:10.1021/jp907054h}
C.~Kn{\"o}fel, C.~Martin, V.~Hornebecq and P.~L. Llewellyn, \emph{The Journal of Physical Chemistry C}, 2009, \textbf{113}, 21726--21734\relax
\mciteBstWouldAddEndPuncttrue
\mciteSetBstMidEndSepPunct{\mcitedefaultmidpunct}
{\mcitedefaultendpunct}{\mcitedefaultseppunct}\relax
\EndOfBibitem
\bibitem[Di~Giovanni \emph{et~al.}(2001)Di~Giovanni, D{\"o}rfler, Mazzotti, and Morbidelli]{doi:10.1021/la010061q}
O.~Di~Giovanni, W.~D{\"o}rfler, M.~Mazzotti and M.~Morbidelli, \emph{Langmuir}, 2001, \textbf{17}, 4316--4321\relax
\mciteBstWouldAddEndPuncttrue
\mciteSetBstMidEndSepPunct{\mcitedefaultmidpunct}
{\mcitedefaultendpunct}{\mcitedefaultseppunct}\relax
\EndOfBibitem
\bibitem[Shi \emph{et~al.}(2024)Shi, Zhang, Xie, Wei, Gong, and Sun]{shi2024characterizing}
J.~Shi, T.~Zhang, X.~Xie, W.~Wei, L.~Gong and S.~Sun, \emph{Computational Geosciences}, 2024,  1--11\relax
\mciteBstWouldAddEndPuncttrue
\mciteSetBstMidEndSepPunct{\mcitedefaultmidpunct}
{\mcitedefaultendpunct}{\mcitedefaultseppunct}\relax
\EndOfBibitem
\bibitem[Mi \emph{et~al.}(2024)Mi, Zhang, and Ge]{doi:10.1021/acs.langmuir.4c03177}
S.~Mi, Y.~Zhang and W.~Ge, \emph{Langmuir}, 2024, \textbf{40}, 21855--21865\relax
\mciteBstWouldAddEndPuncttrue
\mciteSetBstMidEndSepPunct{\mcitedefaultmidpunct}
{\mcitedefaultendpunct}{\mcitedefaultseppunct}\relax
\EndOfBibitem
\bibitem[Liu \emph{et~al.}(2024)Liu, Zhang, Pang, and Sun]{liu2024thermal}
J.~Liu, T.~Zhang, S.~Pang and S.~Sun, \emph{Applied Thermal Engineering}, 2024, \textbf{254}, 123920\relax
\mciteBstWouldAddEndPuncttrue
\mciteSetBstMidEndSepPunct{\mcitedefaultmidpunct}
{\mcitedefaultendpunct}{\mcitedefaultseppunct}\relax
\EndOfBibitem
\bibitem[Khosrowshahi \emph{et~al.}(2022)Khosrowshahi, Abdol, Mashhadimoslem, Khakpour, Emrooz, Sadeghzadeh, and Ghaemi]{khosrowshahi2022role}
M.~S. Khosrowshahi, M.~A. Abdol, H.~Mashhadimoslem, E.~Khakpour, H.~B.~M. Emrooz, S.~Sadeghzadeh and A.~Ghaemi, \emph{Scientific Reports}, 2022, \textbf{12}, 8917\relax
\mciteBstWouldAddEndPuncttrue
\mciteSetBstMidEndSepPunct{\mcitedefaultmidpunct}
{\mcitedefaultendpunct}{\mcitedefaultseppunct}\relax
\EndOfBibitem
\bibitem[Gong \emph{et~al.}(2020)Gong, Shi, Ding, Huang, Sun, and Yao]{GONG2020118406}
L.~Gong, J.-H. Shi, B.~Ding, Z.-Q. Huang, S.-Y. Sun and J.~Yao, \emph{Fuel}, 2020, \textbf{278}, 118406\relax
\mciteBstWouldAddEndPuncttrue
\mciteSetBstMidEndSepPunct{\mcitedefaultmidpunct}
{\mcitedefaultendpunct}{\mcitedefaultseppunct}\relax
\EndOfBibitem
\bibitem[Shi \emph{et~al.}(2019)Shi, Gong, Sun, Huang, Ding, and Yao]{C9RA04963K}
J.~Shi, L.~Gong, S.~Sun, Z.~Huang, B.~Ding and J.~Yao, \emph{RSC Adv.}, 2019, \textbf{9}, 25326--25335\relax
\mciteBstWouldAddEndPuncttrue
\mciteSetBstMidEndSepPunct{\mcitedefaultmidpunct}
{\mcitedefaultendpunct}{\mcitedefaultseppunct}\relax
\EndOfBibitem
\bibitem[Liu \emph{et~al.}(2022)Liu, Zhang, and Sun]{doi:10.1021/acs.energyfuels.2c03244}
J.~Liu, T.~Zhang and S.~Sun, \emph{Energy \& Fuels}, 2022, \textbf{36}, 14865--14873\relax
\mciteBstWouldAddEndPuncttrue
\mciteSetBstMidEndSepPunct{\mcitedefaultmidpunct}
{\mcitedefaultendpunct}{\mcitedefaultseppunct}\relax
\EndOfBibitem
\bibitem[Yiannourakou \emph{et~al.}(2013)Yiannourakou, Ungerer, Leblanc, Rozanska, Saxe, Vidal-Gilbert, Gouth, and Montel]{yiannourakou2013molecular}
M.~Yiannourakou, P.~Ungerer, B.~Leblanc, X.~Rozanska, P.~Saxe, S.~Vidal-Gilbert, F.~Gouth and F.~Montel, \emph{Oil \& Gas Science and Technology--Revue d’IFP Energies nouvelles}, 2013, \textbf{68}, 977--994\relax
\mciteBstWouldAddEndPuncttrue
\mciteSetBstMidEndSepPunct{\mcitedefaultmidpunct}
{\mcitedefaultendpunct}{\mcitedefaultseppunct}\relax
\EndOfBibitem
\bibitem[Turchi \emph{et~al.}(2024)Turchi, Galmarini, and Lunati]{TURCHI2024122709}
M.~Turchi, S.~Galmarini and I.~Lunati, \emph{Journal of Non-Crystalline Solids}, 2024, \textbf{624}, 122709\relax
\mciteBstWouldAddEndPuncttrue
\mciteSetBstMidEndSepPunct{\mcitedefaultmidpunct}
{\mcitedefaultendpunct}{\mcitedefaultseppunct}\relax
\EndOfBibitem
\bibitem[Cygan \emph{et~al.}(2004)Cygan, Liang, and Kalinichev]{doi:10.1021/jp0363287}
R.~T. Cygan, J.-J. Liang and A.~G. Kalinichev, \emph{The Journal of Physical Chemistry B}, 2004, \textbf{108}, 1255--1266\relax
\mciteBstWouldAddEndPuncttrue
\mciteSetBstMidEndSepPunct{\mcitedefaultmidpunct}
{\mcitedefaultendpunct}{\mcitedefaultseppunct}\relax
\EndOfBibitem
\bibitem[Wang \emph{et~al.}(2018)Wang, Zhang, Han, and Weinan]{wang2018deepmd}
H.~Wang, L.~Zhang, J.~Han and E.~Weinan, \emph{Computer Physics Communications}, 2018, \textbf{228}, 178--184\relax
\mciteBstWouldAddEndPuncttrue
\mciteSetBstMidEndSepPunct{\mcitedefaultmidpunct}
{\mcitedefaultendpunct}{\mcitedefaultseppunct}\relax
\EndOfBibitem
\bibitem[Sui \emph{et~al.}(2024)Sui, Zhang, Zhang, Wang, Wang, Yang, and Yao]{sui2024competitive}
H.~Sui, F.~Zhang, L.~Zhang, D.~Wang, Y.~Wang, Y.~Yang and J.~Yao, \emph{Science of The Total Environment}, 2024, \textbf{908}, 168356\relax
\mciteBstWouldAddEndPuncttrue
\mciteSetBstMidEndSepPunct{\mcitedefaultmidpunct}
{\mcitedefaultendpunct}{\mcitedefaultseppunct}\relax
\EndOfBibitem
\bibitem[Deng \emph{et~al.}(2024)Deng, Fang, Liu, Li, Xu, and Chen]{deng2024mineralization}
X.~Deng, N.~Fang, X.~Liu, M.~Li, P.~Xu and Z.~Chen, \emph{Energy \& Fuels}, 2024, \textbf{38}, 12777--12790\relax
\mciteBstWouldAddEndPuncttrue
\mciteSetBstMidEndSepPunct{\mcitedefaultmidpunct}
{\mcitedefaultendpunct}{\mcitedefaultseppunct}\relax
\EndOfBibitem
\bibitem[Stallons and Iglesia(2001)]{STALLONS20014205}
J.~M. Stallons and E.~Iglesia, \emph{Chemical Engineering Science}, 2001, \textbf{56}, 4205--4216\relax
\mciteBstWouldAddEndPuncttrue
\mciteSetBstMidEndSepPunct{\mcitedefaultmidpunct}
{\mcitedefaultendpunct}{\mcitedefaultseppunct}\relax
\EndOfBibitem
\bibitem[Manokaran \emph{et~al.}(2024)Manokaran, Farrusseng, and Coasne]{doi:10.1021/acs.langmuir.4c02136}
R.~Manokaran, D.~Farrusseng and B.~Coasne, \emph{Langmuir}, 2024, \textbf{40}, 22027--22036\relax
\mciteBstWouldAddEndPuncttrue
\mciteSetBstMidEndSepPunct{\mcitedefaultmidpunct}
{\mcitedefaultendpunct}{\mcitedefaultseppunct}\relax
\EndOfBibitem
\bibitem[Long \emph{et~al.}(2021)Long, Lin, Yan, Bai, Tong, Kong, and Li]{long2021adsorption}
H.~Long, H.-f. Lin, M.~Yan, Y.~Bai, X.~Tong, X.-g. Kong and S.-g. Li, \emph{Fuel}, 2021, \textbf{292}, 120268\relax
\mciteBstWouldAddEndPuncttrue
\mciteSetBstMidEndSepPunct{\mcitedefaultmidpunct}
{\mcitedefaultendpunct}{\mcitedefaultseppunct}\relax
\EndOfBibitem
\bibitem[Liu and Bhatia(2013)]{liu2013molecular}
L.~Liu and S.~K. Bhatia, \emph{The Journal of Physical Chemistry C}, 2013, \textbf{117}, 13479--13491\relax
\mciteBstWouldAddEndPuncttrue
\mciteSetBstMidEndSepPunct{\mcitedefaultmidpunct}
{\mcitedefaultendpunct}{\mcitedefaultseppunct}\relax
\EndOfBibitem
\bibitem[Vashishta \emph{et~al.}(1990)Vashishta, Kalia, Rino, and Ebbsj\"o]{PhysRevB.41.12197}
P.~Vashishta, R.~K. Kalia, J.~P. Rino and I.~Ebbsj\"o, \emph{Phys. Rev. B}, 1990, \textbf{41}, 12197--12209\relax
\mciteBstWouldAddEndPuncttrue
\mciteSetBstMidEndSepPunct{\mcitedefaultmidpunct}
{\mcitedefaultendpunct}{\mcitedefaultseppunct}\relax
\EndOfBibitem
\bibitem[Emami \emph{et~al.}(2014)Emami, Puddu, Berry, Varshney, Patwardhan, Perry, and Heinz]{doi:10.1021/cm500365c}
F.~S. Emami, V.~Puddu, R.~J. Berry, V.~Varshney, S.~V. Patwardhan, C.~C. Perry and H.~Heinz, \emph{Chemistry of Materials}, 2014, \textbf{26}, 2647--2658\relax
\mciteBstWouldAddEndPuncttrue
\mciteSetBstMidEndSepPunct{\mcitedefaultmidpunct}
{\mcitedefaultendpunct}{\mcitedefaultseppunct}\relax
\EndOfBibitem
\bibitem[Plimpton(1995)]{plimpton1995fast}
S.~Plimpton, \emph{Journal of Computational Physics}, 1995, \textbf{117}, 1--19\relax
\mciteBstWouldAddEndPuncttrue
\mciteSetBstMidEndSepPunct{\mcitedefaultmidpunct}
{\mcitedefaultendpunct}{\mcitedefaultseppunct}\relax
\EndOfBibitem
\bibitem[Van~Duin \emph{et~al.}(2001)Van~Duin, Dasgupta, Lorant, and Goddard]{van2001reaxff}
A.~C. Van~Duin, S.~Dasgupta, F.~Lorant and W.~A. Goddard, \emph{The Journal of Physical Chemistry A}, 2001, \textbf{105}, 9396--9409\relax
\mciteBstWouldAddEndPuncttrue
\mciteSetBstMidEndSepPunct{\mcitedefaultmidpunct}
{\mcitedefaultendpunct}{\mcitedefaultseppunct}\relax
\EndOfBibitem
\bibitem[Zou and Van~Duin(2012)]{zou2012investigation}
C.~Zou and A.~Van~Duin, \emph{Jom}, 2012, \textbf{64}, 1426--1437\relax
\mciteBstWouldAddEndPuncttrue
\mciteSetBstMidEndSepPunct{\mcitedefaultmidpunct}
{\mcitedefaultendpunct}{\mcitedefaultseppunct}\relax
\EndOfBibitem
\bibitem[Abascal and Vega(2005)]{abascal2005general}
J.~L.~F. Abascal and C.~Vega, \emph{The Journal of Chemical Physics}, 2005, \textbf{123}, 234505\relax
\mciteBstWouldAddEndPuncttrue
\mciteSetBstMidEndSepPunct{\mcitedefaultmidpunct}
{\mcitedefaultendpunct}{\mcitedefaultseppunct}\relax
\EndOfBibitem
\bibitem[Potoff and Siepmann(2001)]{potoff2001vapor}
J.~J. Potoff and J.~I. Siepmann, \emph{AIChE journal}, 2001, \textbf{47}, 1676--1682\relax
\mciteBstWouldAddEndPuncttrue
\mciteSetBstMidEndSepPunct{\mcitedefaultmidpunct}
{\mcitedefaultendpunct}{\mcitedefaultseppunct}\relax
\EndOfBibitem
\bibitem[Wisniak(2010)]{wisniak2010daniel}
J.~Wisniak, \emph{Educaci{\'o}n qu{\'\i}mica}, 2010, \textbf{21}, 155--162\relax
\mciteBstWouldAddEndPuncttrue
\mciteSetBstMidEndSepPunct{\mcitedefaultmidpunct}
{\mcitedefaultendpunct}{\mcitedefaultseppunct}\relax
\EndOfBibitem
\bibitem[K{\"u}hne \emph{et~al.}(2020)K{\"u}hne, Iannuzzi, Del~Ben, Rybkin, Seewald, Stein, Laino, Khaliullin, Sch{\"u}tt, Schiffmann,\emph{et~al.}]{kuhne2020cp2k}
T.~D. K{\"u}hne, M.~Iannuzzi, M.~Del~Ben, V.~V. Rybkin, P.~Seewald, F.~Stein, T.~Laino, R.~Z. Khaliullin, O.~Sch{\"u}tt, F.~Schiffmann \emph{et~al.}, \emph{The Journal of Chemical Physics}, 2020, \textbf{152}, 194103\relax
\mciteBstWouldAddEndPuncttrue
\mciteSetBstMidEndSepPunct{\mcitedefaultmidpunct}
{\mcitedefaultendpunct}{\mcitedefaultseppunct}\relax
\EndOfBibitem
\bibitem[VandeVondele and Hutter(2003)]{vandevondele2003efficient}
J.~VandeVondele and J.~Hutter, \emph{The Journal of chemical physics}, 2003, \textbf{118}, 4365--4369\relax
\mciteBstWouldAddEndPuncttrue
\mciteSetBstMidEndSepPunct{\mcitedefaultmidpunct}
{\mcitedefaultendpunct}{\mcitedefaultseppunct}\relax
\EndOfBibitem
\bibitem[Lippert \emph{et~al.}(1999)Lippert, Hutter, and Parrinello]{lippert1999gaussian}
G.~Lippert, J.~Hutter and M.~Parrinello, \emph{Theoretical Chemistry Accounts}, 1999, \textbf{103}, 124--140\relax
\mciteBstWouldAddEndPuncttrue
\mciteSetBstMidEndSepPunct{\mcitedefaultmidpunct}
{\mcitedefaultendpunct}{\mcitedefaultseppunct}\relax
\EndOfBibitem
\bibitem[VandeVondele and Hutter(2007)]{vandevondele2007gaussian}
J.~VandeVondele and J.~Hutter, \emph{The Journal of Chemical Physics}, 2007, \textbf{127}, 114105\relax
\mciteBstWouldAddEndPuncttrue
\mciteSetBstMidEndSepPunct{\mcitedefaultmidpunct}
{\mcitedefaultendpunct}{\mcitedefaultseppunct}\relax
\EndOfBibitem
\bibitem[Perdew \emph{et~al.}(1996)Perdew, Burke, and Ernzerhof]{perdew1996generalized}
J.~P. Perdew, K.~Burke and M.~Ernzerhof, \emph{Physical review letters}, 1996, \textbf{77}, 3865\relax
\mciteBstWouldAddEndPuncttrue
\mciteSetBstMidEndSepPunct{\mcitedefaultmidpunct}
{\mcitedefaultendpunct}{\mcitedefaultseppunct}\relax
\EndOfBibitem
\bibitem[Grimme(2006)]{grimme2006semiempirical}
S.~Grimme, \emph{Journal of computational chemistry}, 2006, \textbf{27}, 1787--1799\relax
\mciteBstWouldAddEndPuncttrue
\mciteSetBstMidEndSepPunct{\mcitedefaultmidpunct}
{\mcitedefaultendpunct}{\mcitedefaultseppunct}\relax
\EndOfBibitem
\bibitem[Knight \emph{et~al.}(2012)Knight, Lindberg, and Voth]{knight2012multiscale}
C.~Knight, G.~E. Lindberg and G.~A. Voth, \emph{The Journal of Chemical Physics}, 2012, \textbf{137}, 22A525\relax
\mciteBstWouldAddEndPuncttrue
\mciteSetBstMidEndSepPunct{\mcitedefaultmidpunct}
{\mcitedefaultendpunct}{\mcitedefaultseppunct}\relax
\EndOfBibitem
\bibitem[Goedecker \emph{et~al.}(1996)Goedecker, Teter, and Hutter]{goedecker1996separable}
S.~Goedecker, M.~Teter and J.~Hutter, \emph{Physical Review B}, 1996, \textbf{54}, 1703\relax
\mciteBstWouldAddEndPuncttrue
\mciteSetBstMidEndSepPunct{\mcitedefaultmidpunct}
{\mcitedefaultendpunct}{\mcitedefaultseppunct}\relax
\EndOfBibitem
\bibitem[Peng and Robinson(1976)]{peng1976new}
D.-Y. Peng and D.~B. Robinson, \emph{Industrial \& Engineering Chemistry Fundamentals}, 1976, \textbf{15}, 59--64\relax
\mciteBstWouldAddEndPuncttrue
\mciteSetBstMidEndSepPunct{\mcitedefaultmidpunct}
{\mcitedefaultendpunct}{\mcitedefaultseppunct}\relax
\EndOfBibitem
\bibitem[Gomaa \emph{et~al.}(2022)Gomaa, Guerrero, Heidari, and Espinoza]{gomaa2022experimental}
I.~Gomaa, J.~Guerrero, Z.~Heidari and D.~N. Espinoza, SPE Annual Technical Conference and Exhibition?, 2022, p. D011S018R002\relax
\mciteBstWouldAddEndPuncttrue
\mciteSetBstMidEndSepPunct{\mcitedefaultmidpunct}
{\mcitedefaultendpunct}{\mcitedefaultseppunct}\relax
\EndOfBibitem
\bibitem[{\c{S}}ahin \emph{et~al.}(2022){\c{S}}ahin, Arslan, and Tomul]{csahin2022adsorption}
M.~{\c{S}}ahin, Y.~Arslan and F.~Tomul, \emph{International Journal of Environmental Analytical Chemistry}, 2022,  1--20\relax
\mciteBstWouldAddEndPuncttrue
\mciteSetBstMidEndSepPunct{\mcitedefaultmidpunct}
{\mcitedefaultendpunct}{\mcitedefaultseppunct}\relax
\EndOfBibitem
\bibitem[Narasimharao \emph{et~al.}(2022)Narasimharao, Al-Thabaiti, Rajor, Mokhtar, Alsheshri, Alfaifi, Siddiqui, and Abdulla]{narasimharao2022fe3o4}
K.~Narasimharao, S.~Al-Thabaiti, H.~K. Rajor, M.~Mokhtar, A.~Alsheshri, S.~Y. Alfaifi, S.~I. Siddiqui and N.~K. Abdulla, \emph{Journal of Materials Research and Technology}, 2022, \textbf{18}, 3581--3597\relax
\mciteBstWouldAddEndPuncttrue
\mciteSetBstMidEndSepPunct{\mcitedefaultmidpunct}
{\mcitedefaultendpunct}{\mcitedefaultseppunct}\relax
\EndOfBibitem
\bibitem[Shafeeyan \emph{et~al.}(2015)Shafeeyan, Daud, Shamiri, and Aghamohammadi]{shafeeyan2015modeling}
M.~S. Shafeeyan, W.~M. A.~W. Daud, A.~Shamiri and N.~Aghamohammadi, \emph{Energy \& Fuels}, 2015, \textbf{29}, 6565--6577\relax
\mciteBstWouldAddEndPuncttrue
\mciteSetBstMidEndSepPunct{\mcitedefaultmidpunct}
{\mcitedefaultendpunct}{\mcitedefaultseppunct}\relax
\EndOfBibitem
\bibitem[Marczewski(2010)]{marczewski2010application}
A.~Marczewski, \emph{Applied Surface Science}, 2010, \textbf{256}, 5145--5152\relax
\mciteBstWouldAddEndPuncttrue
\mciteSetBstMidEndSepPunct{\mcitedefaultmidpunct}
{\mcitedefaultendpunct}{\mcitedefaultseppunct}\relax
\EndOfBibitem
\bibitem[Shi \emph{et~al.}(2019)Shi, Gong, Sun, Huang, Ding, and Yao]{shi2019competitive}
J.~Shi, L.~Gong, S.~Sun, Z.~Huang, B.~Ding and J.~Yao, \emph{Rsc Advances}, 2019, \textbf{9}, 25326--25335\relax
\mciteBstWouldAddEndPuncttrue
\mciteSetBstMidEndSepPunct{\mcitedefaultmidpunct}
{\mcitedefaultendpunct}{\mcitedefaultseppunct}\relax
\EndOfBibitem
\bibitem[Izvekov and Voth(2005)]{izvekov2005ab}
S.~Izvekov and G.~A. Voth, \emph{The Journal of Chemical Physics}, 2005, \textbf{123}, 044505\relax
\mciteBstWouldAddEndPuncttrue
\mciteSetBstMidEndSepPunct{\mcitedefaultmidpunct}
{\mcitedefaultendpunct}{\mcitedefaultseppunct}\relax
\EndOfBibitem
\bibitem[Zundel(1969)]{zundel1969hydration}
G.~Zundel, \emph{Angewandte Chemie International Edition in English}, 1969, \textbf{8}, 499--509\relax
\mciteBstWouldAddEndPuncttrue
\mciteSetBstMidEndSepPunct{\mcitedefaultmidpunct}
{\mcitedefaultendpunct}{\mcitedefaultseppunct}\relax
\EndOfBibitem
\bibitem[Agmon(1995)]{agmon1995grotthuss}
N.~Agmon, \emph{Chemical Physics Letters}, 1995, \textbf{244}, 456--462\relax
\mciteBstWouldAddEndPuncttrue
\mciteSetBstMidEndSepPunct{\mcitedefaultmidpunct}
{\mcitedefaultendpunct}{\mcitedefaultseppunct}\relax
\EndOfBibitem
\bibitem[Gomaa \emph{et~al.}(2023)Gomaa, Guerrero, Heidari, and Espinoza]{gomaa2023experimental}
I.~Gomaa, J.~Guerrero, Z.~Heidari and D.~N. Espinoza, \emph{SPE Reservoir Evaluation \& Engineering}, 2023,  1--14\relax
\mciteBstWouldAddEndPuncttrue
\mciteSetBstMidEndSepPunct{\mcitedefaultmidpunct}
{\mcitedefaultendpunct}{\mcitedefaultseppunct}\relax
\EndOfBibitem
\end{mcitethebibliography}
\bibliographystyle{rsc} %the RSC's .bst file

\end{document}